\documentclass[twocolumn,english,reprint,longbibliography,superscriptaddress,aip]{revtex4-1}
\usepackage[utf8]{inputenc}
\usepackage{amsmath,amssymb}
\usepackage{xcolor}
\usepackage{graphicx}
\usepackage{dcolumn}
\usepackage{bm}
\usepackage{comment}
\usepackage[normalem]{ulem}
\usepackage{lipsum}
\usepackage{color}
\usepackage{xcolor}
\usepackage{simplewick}
\usepackage{qcircuit}
\usepackage{braket}
\usepackage{float}
\usepackage{multirow}
\usepackage{tikz}
\usepackage{multibib}
\usepackage{appendix}
\usepackage[colorlinks=true,%
bookmarks=false,%
linkcolor=black,%
urlcolor=blue,%
citecolor=blue,%
breaklinks]{hyperref}
\usepackage{soul}

\usepackage{subfig}

\makeatother

\begin{document}

\author{Guglielmo Mazzola}
\affiliation{Institute for Computational Science, University of Zurich, Winterthurerstrasse 190, CH-8057 Zurich, Switzerland}

\date{today}

\title{Quantum computing for chemistry and physics applications\\  from a Monte Carlo perspective}
\date{\today}

\begin{abstract}
This Perspective focuses on the several overlaps between quantum algorithms and Monte Carlo methods in the domains of physics and chemistry. We will analyze the challenges and possibilities of integrating established quantum Monte Carlo solutions in quantum algorithms. These include refined energy estimators, parameter optimization, real and imaginary-time dynamics, and variational circuits.
Conversely, we will review new ideas in utilizing quantum hardware to accelerate the sampling in statistical classical models, with applications in physics, chemistry, optimization, and machine learning.
This review aims to be accessible to both communities and intends to foster further algorithmic developments  at the intersection of quantum computing and Monte Carlo methods.
Most of the works discussed in this Perspective have emerged within the last two years, indicating a rapidly growing interest in this promising area of research.

\end{abstract}

\maketitle

\tableofcontents

\section{Introduction} 

The solution of quantum many-body problems in chemistry and physics is one of the most anticipated applications of a quantum computer, as first proposed by Feynman.\cite{feynman_simulating_1982}
Over time, it has been proposed that many other classes of problems can benefit from quantum speedup, including cryptography, data science, machine learning, finance, linear algebra, and optimization.\cite{ichikawa2023comprehensive} However, physics and chemistry remain among the main candidates for demonstrating practical quantum advantage over conventional methods because they contain classes of problems with the following characteristics: (i) they are very challenging for classical computation, and exponential quantum speed-ups are possible, and (ii) they are defined by a small number of variables, thus featuring a limited cost of data loading and reading.\cite{hoefler2023disentangling}

Among all possible problems in physics, here we will focus on electronic structure and spin models (including classical spin models), as their implementation requires a relatively lower cost compared to models like high-energy physics.
Excellent review articles on quantum algorithms for quantum chemistry\cite{Cao_2019,mcardle2018quantum} and materials science\cite{bauer2020quantum} have already been published a few years ago,  the latest one in 2020. The purpose of this manuscript is not to duplicate such presentations but rather to concentrate on a more frontier topic that is becoming relevant due to several works appearing in recent months.
However, notice that about $40 \%$ of references cited in this Perspective are pretty recent, i.e. from 2021 onwards. This indicates how fast the whole field of quantum computing is growing.

We will analyze points of contact between quantum computing and Monte Carlo (MC), quantum Monte Carlo (QMC) methods.\cite{becca_sorella_2017}
There are many common themes between the two worlds. Shot noise arising from the measurements of the quantum register finds a parallel in the statistical root of Monte Carlo. Both methods require extracting and utilizing expectation values computed in the presence of statistical noise. 

The existence of shot noise is one of the major issues for near-term simulations: in variational setting this implies a problematically large number of circuit repetitions.\cite{wecker2015progress} On the other hand, such uncorrelated wave function collapses can have computational value if used as importance sampling in Monte Carlo.
We will also report attempts of cross-fertilization between the two fields in designing variational ansatze and optimization methods for ground state and dynamical problems.
Additionally, we discuss the requirements that a classical-quantum hybrid QMC algorithm, relying on a quantum computing subroutine, must meet. Regarding classical applications, we review several proposals for accelerating the Metropolis algorithm using quantum hardware and examine their practicality under realistic hardware constraints.

Therefore, the purpose of this manuscript is to review various Monte Carlo techniques that can be useful for creating new quantum algorithms or designing new applications of already-known quantum primitives.
Conversely, this Perspective also aims to be an accessible presentation of the potential and limitations of quantum computing, for  Monte Carlo experts and, more broadly, computational physicists.

\begin{figure*}[ht!]
\includegraphics[width=0.8\textwidth]{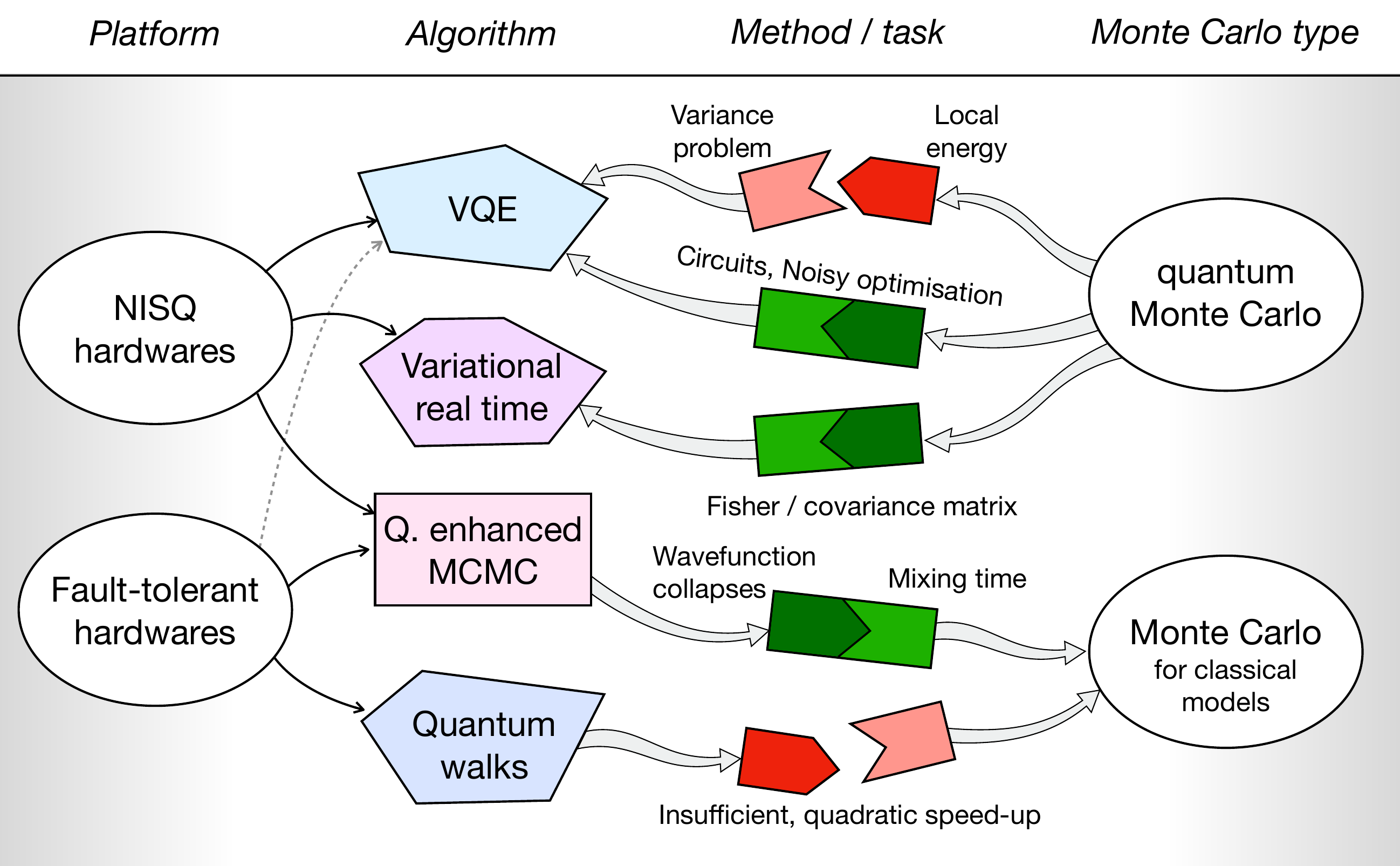}
\caption{Map of the main links, between quantum algorithms and Monte Carlo methods, contained in this Perspective. 
Connected green links indicate that fruitful information flow between the two area has already been established. Disconnected red links indicate topics that still require more investigation or where proposed solutions are not completely satisfactory.
}
\label{fig:map}
\end{figure*} 

This Perspective is timely as (1) 
on the experimental side, the first steps toward fault-tolerant hardware have been made. \cite{krinner2022realizing,sivak2023real,sivak2022real,postler2022demonstration} Moreover, experiments at the threshold of quantum advantage for quantum dynamics are now possible given the existence of $\sim 100$ qubits hardwares, albeit noisy.\cite{kim2023evidence}
(2) On the quantum  variational algorithm side, we have increasing evidence that the quantum measurement noise -the focus of this Perspective- is the major, unavoidable, bottleneck of near-term quantum algorithms.\cite{wecker2015progress,astrakhantsev2022algorithmic}
(3) In the last year, thorough resource assessment papers for quantum chemistry have appeared,\cite{vonBurg2021catalysis,goings2022reliably,beverland2022assessing} which clearly reaffirms the threshold for quantum advantage for ground state problems, at least with today's algorithms, to be deep into the future fault-tolerant quantum computing regime, and question the previously claimed exponential advantage for ground-state electronic structure application.\cite{lee2022there}
(4) Finally, we observe the emergence of a new class of hybrid quantum algorithms revisiting classical
and quantum Monte Carlo, opening completely new possibilities for quantum advantage in these areas.
In this Perspective we report about twenty recent (i.e which appeared in the last two years) works that aim to complement quantum computing and Monte Carlo in several sub-fields: hybrid quantum computing-QMC, variational circuits development, parameter optimization, time-dependent simulations, and classical sampling.

The manuscript is organized as follows. In Sec.~\ref{s:hardware} we briefly mention the types of quantum hardware and their fundamental limitation, namely the hardware errors in the NISQ regime, and the fairly long (compared to conventional CPUs) gate time of future, error-corrected machines. In Sec.~\ref{s:algo} we introduce general quantum algorithms for physics and the requirements for quantum advantage. In Sec.~\ref{s:ferm} we review the basics of the encoding of a fermionic hamiltonian into a quantum computer.
After these introductions, in Sec.~\ref{s:variational}, we discuss the variational method, the simplest kind of algorithm for physics and chemistry, and its limitation due to the shot noise.
In Sec.~\ref{s:fault_tole} we instead describe how the same calculation could be done in the fault-tolerant era.
The more technical Sec.~\ref{sect:vmc} introduces the local energy estimator that is central in QMC and explains why these algorithms, while still being stochastic, do not suffer from the severe variance problem of variational quantum algorithms. 
In Sec.~\ref{sect:qcqmc} we review attempts to create hybrid quantum-classical QMC algorithms, as well as other points of contact between QMC and quantum algorithms. 
Finally, in Sec.~\ref{sect:sampling}, we reverse our perspective and discuss quantum computing methods to speed-up classical sampling, using digital machines and quantum simulators, annealers.

The concept map of the Perspective is shown in Fig.~\ref{fig:map}.

\section{Quantum hardwares}
\label{s:hardware}

Unlike conventional reviews on algorithms for quantum chemistry, it is necessary to briefly introduce the hardware on which these must be executed. Understanding the possibilities and limitations of the hardware is crucial to get an idea of the feasibility of current and future algorithms.

There are many types of quantum computers and quantum simulators. The difference between the two classes is that a quantum computer is built with the idea of being universal, therefore able to support any type of program. A quantum simulator is designed to perform a narrower range of tasks, such as optimizing classical cost-functions,\cite{johnson2011quantum,pagano2020quantum} or simulating particular  Hamiltonians.\cite{RevModPhys.93.025001}

To tend towards universality, a quantum computer must support the execution of basic local operations, called \emph{quantum gates}, just like a regular computer. For example, an architecture capable of synthesizing the following set of gates $\{ \textsf{CNOT}, \textsf{H}, \textsf{S} , \textsf{T} \}$ or  $\{ \textsf{R}_x(\theta), \textsf{R}_y(\theta), \textsf{R}_z(\theta),  \textsf{S} , \textsf{CNOT} \}$ that act on a maximum of two qubits (see textbooks on quantum computing for the definition of these gates\cite{nielsen_chuang_2010}), is capable of approximating any possible unitary operation on $n$ qubits.\cite{nielsen_chuang_2010} On the other hand, a special-purpose quantum simulator can implement the global operation of interest directly, such as $e^{iHt}$, where $H$ is the quantum many-body Hamiltonian operator, without the problem of having to compile it using a gate set.\cite{daley_practical_2022}
Currently, the greatest engineering effort is focusing on  gate-based ``digital quantum computers'', although it is not excluded that algorithms of interest for chemistry and materials science can be executed on quantum simulators.

There is then a second distinction that is important to keep in mind. At present, the size of digital quantum computers is on the order of $n \approx 100$  qubits.\cite{arute:2019,kim2023evidence} In principle, these computers have access to a $2^{\mathcal{O}(100)}$ dimensional Hilbert space. However, practical quantum advantage has not been achieved yet. The reason is that these machines are not properly digital ones but are subject to hardware noise. For example, each gate has a finite insuccess probability. As we will see, circuits necessary to write a quantum algorithm for chemistry require a considerable number of gates, and therefore even a small infidelity propagates devastatingly and the total error accumulates until it completely compromises the success of the algorithm. Current hardwares are built with the idea of executing a universal gate set, but are still affected by hardware noise. They generally go by the name of NISQ (Noisy Intermediate-Scale Quantum) machines.\cite{preskill2018quantum}

The final step to achieving a true digital quantum computer is to realize hardwares capable of executing gates without errors, just like their classical counterpart. Many detractors of quantum computers base their skepticism on the impossibility of maintaining a macroscopic coherent wavefunction for an arbitrary number of operations.\cite{laughlin2005reinventing} Fortunately, there is a theorem that does not exclude the possibility of a digital universal computer: below a certain noise threshold, it is possible to correct this hardware noise faster than it can accumulate during runtime (the same would not hold for analog computers).\cite{nielsen_chuang_2010} Practical proposals to realize this idea include using multiple physical qubits to realize a logical qubit and more operations to realize a single ``error-corrected'' logical gate.\cite{fowler2012surface,litinski2019game}

It is important to understand which types of algorithms have the hope of being executed on a NISQ machine and which will require a fault-tolerant machine to properly contextualize the ever-growing literature on quantum computing for quantum chemistry. At present, sub-communities have formed that are dedicated to developing NISQ algorithms, and others, that are increasingly growing in number, which are developing algorithms for the fault-tolerant era.
This situation is unprecedented. In classical computing, to attempt a comparison, it would be as if in the 1960's there were a community developing algorithms for chemistry on punched cards, and another preparing for exascale computing without a clear idea of whether and how an HPC facility would be built.

However, the technological progress toward a fault-tolerant machine is steady. 
Several experiments published in 2022-2023 already demonstrated some building blocks necessary for quantum error correction.\cite{PhysRevX.11.041058,krinner2022realizing,sivak2022real,postler2022demonstration,sundaresan2023demonstrating,google2023suppressing}
Clearly, we are still in the infancy of fault-tolerant hardware, and it is not yet clear when a large-scale error-corrected machine, able to accommodate electronic structure calculations, will appear. This also depends on progress in the algorithmic side of compressing memory and runtime resources.

The last concept that needs to be presented in this brief account of hardware is related to the clock frequency of a quantum computer, which will always be necessarily slower than classical gates. This is because the execution of a quantum gate requires the manipulation of a wave function by an external control: the quantum gate can never be faster than the classical electronic apparatus that controls it.
At present, the execution time of a noisy CNOT or noisy single qubit rotation is of the order of 100 ns in the NISQ era, for superconducting qubits, corresponding to 100 MHz.\cite{arute:2019} The expected logical clock rate in the fault-tolerant regime is much slower, of the order of 10-100kHz, because every error-corrected gate requires a large number of elementary operations that involve a large number of physical qubits.\cite{gidney2019efficient}
Often, in literature, one hears about \textsf{T}-depth as a proxy for the complexity of an algorithm.\cite{Reiher_2017,Chakrabarti2021thresholdquantum,beverland2022assessing} The reason is that truly digital hardware can only operate using a set of discrete gates that can be error-corrected. A rotation of an arbitrary angle does not belong to this category, and therefore every continuous rotation must be compiled into a series of discrete operations, such as \textsf{S}, \textsf{H}, and \textsf{T} gates. The \textsf{T} gates are the most expensive to synthesize (i.e 100-10000 more costly than a \textsf{CNOT}),\cite{gidney2019efficient} and therefore their number, and how many can be executed in parallel, determines the runtime.
Since in chemistry, rotations of an arbitrary angle may be necessary for orbital basis rotations, or to realize infinitesimal Trotter step operations, are ubiquitous, fault-tolerant quantum algorithms generally include a large number of \textsf{T} gates. It is interesting to note that in the NISQ era, the opposite is true: rotations are comparatively simple gates to perform, while efforts are made to reduce the number of \textsf{CNOT} gates, which are currently the most noisy.
Recent proposals include the possibility of retaining analog rotation gates and error-corrected Clifford gates, which are easier to synthetize.\cite{PhysRevLett.127.200505,akahoshi2023partially}
This hybrid approach is interesting but yet to be demonstrated in practical algorithms.
Moreover, the gate times depend on the specific hardware architecture. Other platforms such as spin qubits, trapped ions, or photonic hardwares will imply different hardware constraints.

\section{Algorithms and quantum advantage}
\label{s:algo}

After this essential overview of hardware, we are now in a position to introduce more concretely the most popular algorithms for the quantum many-body problem.
The application where quantum advantage appears most clear and easy to justify is quantum dynamics. In this case, an exponential advantage can be obtained by virtue of the exponential compression of the Hilbert space into a linear memory with the number of particles.\cite{lloyd:1996}
If we consider, for simplicity, spin-1/2 lattice models composed of $n$ spins, it is easy to see that an \emph{exact} simulation becomes unfeasible as soon as $n\sim50$.
Storing a quantum state of 50 qubits with double-precision coefficients for each of the $2^{50}$ possible components requires 16 PB of memory. To perform arbitrary discrete-time evolution, we would need to manipulate such an array thousands of times.
For instance, direct matrix exponentiation $e^{iHt}$ for a typical many-body quantum Hamiltonian, $H$, and 50 spin-1/2 particles would require an array made of $10^{15}$ entries undergoing a matrix-vector multiplication of size $2^{50} \times 2^{50} = 10^{30}$.

The memory requirement for evolving a system of $N$ electrons in second quantization using $n$ orbitals is the same. Second quantization offers an optimal memory usage if a Fock encoding is used, where each binary string represents a possible configuration of orbital occupation (see Sec.~\ref{s:ferm}).\cite{whitfield2010simulation}
As we will see, the choice of second quantization introduces non-locality in the Hamiltonian, which is translated as a sum of tensor products of Pauli matrices (qubits are not fermionic particles) and therefore requires very long circuits. From an asymptotic point of view, first quantization would be a better choice, as it preserves the locality of interactions at the cost of introducing expensive arithmetic operations to calculate the Coulomb term. In addition, an antisymmetric function in real space must be provided.\cite{kivlichan2017bounding}

Simulating lattice spin models, therefore, appears to be the most obvious choice in the search for a quantum advantage in Hamiltonian simulations. Beverland et. al\cite{beverland2022assessing} shows that a fault-tolerant simulation of a $10 \times 10$ plaquette of the two-dimensional quantum Ising model requires on the order of $10^5-10^6$ physical superconducting qubits, using the digital Trotter algorithm. This example is instructive, as it clearly shows that, although the system requires a priori a memory of $n=100$ logical qubits (1 logical qubit = 1 spin), error correction and extra resources for distilling \textsf{T} gates (often called \textsf{T}-factories) push the total computation of qubits up to a million.
This system sets a lower bound on the resources needed to simulate fermionic systems that are more complex than spin-$1/2$ models, for the same number of particles/spin-orbitals. Simulating lattice models is also possible with quantum simulators, although with calibration errors, and it is therefore likely that there will be competition between the two quantum computing paradigms towards the first simulation, beyond what is possible classically.\cite{daley_practical_2022,miessen2023quantum}

Indeed, it is also possible that the first demonstration of practical quantum advantage in real-time simulations will be achieved before the fault-tolerant regime, for example in simulations of dynamical phase transition.\cite{king2022coherent}
At the time of writing this review, researches at IBM showcased a record-sized real-time dynamics of a 2D heavy-hexagon lattice Ising model using Trotterization, 127 qubits, and error mitigation techniques.\cite{kim2023evidence}
Just ten days later, approximate tensor network simulations achieved the same result,\cite{tindall2023efficient,begusic2023fast} raising the bar once again to declare a quantum advantage, similar to what happened with the first claim of quantum supremacy, on random circuit sampling.\cite{arute:2019,huang2020classical,pan2021simulating}
Demonstrating a definitive quantum advantage in quantum dynamics tasks is a less well-defined goal, as classical limits are still not thoroughly explored. One can expect that increasingly sophisticated classical methods will be adopted to counter the new claims of quantum advantage.
The case of chemistry is different since classical methods are fairly established, and it is much clearer which electronic structure Hamiltonians are beyond the reach of classical computing.

Coming back to our chemistry problems, if the required answer requires a precision that can only be achieved with a long circuit, then we must prepare ourselves for the fact that our algorithm will only be available in the digital era, which could take a decade or more, when a technology capable of controlling millions of qubits will be available. As mentioned, fault-tolerant hardware has a lower clock frequency than a conventional computer,  a non-exponential asymptotic speed-up may not be sufficient to guarantee actually shorter runtimes, for reasonable system sizes.
Babbush et. al.\cite{babbush:2021} recently discuss how a quadratic speed-up is insufficient for practical advantage in many applications.

To conclude, quantum advantage in chemistry problems can be obtained either in several years using a fault-tolerant algorithm with a superquadratic speed-up, or in a heuristic way using NISQ hardwares. In this case, only short-depth algorithms can be used, which often require a classical feedback loop, such as optimization of gate parameters that define the circuit, and repeated executions of the circuit. Variational methods fall exactly into this class of solutions.\cite{mcclean2016theory,cerezo2021variational} We will see in the next chapter how these can also be used and their limitations, especially their relationship with a type of noise that cannot be eliminated even in the fault-tolerant regime.

\section{Fermionic hamiltonians}
\label{s:ferm}

Most applications in chemistry are related to solving the many-electron Schr{\"o}dinger equation. To clearly understand the problem it is necessary to formalize it to some extent.
The field of quantum simulations of electronic structure problems is quite well established, and we refer to reviews\cite{mcardle2018quantum,Cao_2019,bauer2020quantum} for a general introduction and more details.
The most general fermionic hamiltonian reads
\begin{equation}
\label{eq:fermionic_ham}
	H = \sum_{p,q}t_{pq}a^\dag_p a_q + \sum_{p,q,r,s}u_{pqrs}a^\dag_p a^\dag_q a_ra_s,
\end{equation}
where $a_p^\dag$ and $a_p$ are the fermionic creation and destruction operators, which create (annihilate) a particle in the spin-orbital $p$. 

To proceed, the next step is to define an encoding from fermionic Hilbert space to qubit space, so that a base vector of the latter, for example \textsf{1100}, has a unique correspondence in the fermionic space.

In second quantization, this is trivial using Fock states. A bit-string  denotes the occupation numbers of spin-orbitals, in a chosen ordering. For instance, the string \textsf{1100} can represent the Hartree-Fock state of an $H_2$ molecule described with two spatial molecular orbitals. Here, there are two electrons, of opposite spin, occupying the first two spin-orbitals, $ \{\ket{\psi_{g,\uparrow}}, \ket{\psi_{g,\downarrow}} \}$, while leaving empty the two higher energy ones $ \{\ket{\psi_{u,\uparrow}}, \ket{\psi_{u,\downarrow}} \}$.

For ease of notation, we can label this string with its binary index (from left to right, here), $\ket{1100} = \ket 3$. The (doubly) excited state, compatible with the symmetries of the molecule, is $\ket{0011} = \ket{12}$. In the $H_2$ case, the exact ground state (in this small atomic basis) can be expressed as a linear combination of these two strings alone, but in the general case, the ground state of a molecule with $N$ electrons and $n$ spin-orbitals is written as a linear combination,
\begin{equation}
\label{eq:ket}
    \ket{\psi} = \sum_{i=0}^{2^n-1} c_i \ket i~,
\end{equation}
where $c_i$ are complex coefficients, and $\ket i$ is a basis vector, which binary format denotes the occupation number of the spin-orbitals in a chosen ordering.
Although many $c_i$'s are zero due to particle and spin conservation, as is known, an exponentially large number of them remains finite when $N$ and $n$ increase.

The main advantage of a quantum computer is therefore clear. With only $n$ qubits, it is possible to store in memory a wavefunction with $2^n$ complex coefficients.
The fundamental question now is whether it is possible to devise a quantum algorithm with polynomial complexity, capable of manipulating these $2^n$ coefficients to find the ground state of a fermionic problem, or at least with a better approximation than the best classical method.

Before going further, it is necessary to show what kind of Hamiltonian is produced after the fermion-to-qubit mapping is applied.
Since a qubit is not a fermion, the hamiltonian in Eq.~\ref{eq:fermionic_ham} needs to be translated into a  qubit operator.
This is usually done with the Jordan-Wigner transformations,
\begin{equation}
	\begin{aligned}
		a_p &= \frac{1}{2}(X + iY)_p \otimes Z_{p-1}\otimes \dots\otimes  Z_{0},\\
		a_p^\dag &= \frac{1}{2}(X - iY)_p \otimes Z_{p-1}\otimes \dots\otimes  Z_{0},
	\end{aligned}
\end{equation}
where the $\frac{1}{2}(X + iY)_p$ and $\frac{1}{2}(X - iY)_p$ spin \emph{minus} and \emph{plus} operators change the occupation number of the target mode $p$, while the string of $Z$ operators are needed to enforce antisymmetrization. 
Therefore, each fermionic operator in Eq.~\ref{eq:fermionic_ham} translates into a combination of tensor product of Pauli operators.

The full Hamiltonian of Eq.~\ref{eq:fermionic_ham} then takes the general form of
 linear combination of products of single-qubit Pauli operators
\begin{equation}
\label{eq:pauli_ham}
	H = \sum_{j=1}^{N_P} h_j P_j 
\end{equation}
where $h_j$ is a real scalar coefficient, and $N_P$ is of order $\mathcal{O}(n^4)$ since there are $n^4$ terms to transform in Eq.~\ref{eq:fermionic_ham}. Each term $P_j$ in the Hamiltonian is typically referred to as a \emph{Pauli string}, and is a tensor product $P_j =\bigotimes_{i=1}^n \sigma_i^\alpha$ of $n$ Pauli matrices $\sigma^\alpha \in \{ I, Z, X, Y \}$. 
What is important for our discussion is that (1) the coefficient $h_p$ can take very large values (in modulus).
Just to give an idea, $\sum_j |h_j|$ is of the order of 40 Ha, already for a moderate example of a $H_2O$ molecule described in STO-6G basis\cite{wecker2015progress}.
Then (2) the operators $P_j$ can have a number of non-identity gates, which is $\mathcal{O}(n)$ due to the non-locality introduced by the Jordan-Wigner trasformation. This implies that the circuit for real-time dynamics are longer compared to local quantum spin Hamiltonians,\cite{whitfield2010simulation} and when measured the expectation value $\langle P_j \rangle$, is exponentially subsceptible to bit-flip measurement errors\cite{huggins2021efficient}. 
The form of Eq.~{\ref{eq:pauli_ham} is however more general than the Jordan-Wigner workflow, so we assume it as a starting point of our discussion.

\section{Variational quantum algorithms}
\label{s:variational}

In short, variational methods aims to use shallow parametrized circuits that can be optimized to minimize the calculated energy.\cite{peruzzo2014,yuan2019,cerezo2021variational}
The $N_{par}$ variational parameters can be the angles $\theta$'s of rotation gates defined above.
This strategy takes the name of Variational Quantum Eigensolver (VQE) but it is nothing more than a common variational calculation on a quantum computing platform.
First, of all, it is important to notice that even shallow circuits, i.e. featuring constant depth vs $n$, can display quantum advantage, although on quite artificial tasks,\cite{bravyi2018quantum} or cannot be efficiently simulated classically.\cite{farhi2016quantum}
Therefore, variational algorithms are reasonable candidates for quantum advantage in the near term.
As of today, the literature features many small-scale hardware demonstrations, still away from the quantum advantage threshold. The most notable either use heuristic circuits,\cite{kandala_hardware-efficient_2017}, or more structured physically inspired circuit ansatze.\cite{google_ai_quantum_and_collaborators_hartree-fock_2020, stanisic2022observing}
The current largest variational simulation of a chemical system reaches a system size of about 20 qubits.\cite{o2022purification}
Performing variational calculations of many-body quantum systems has advantages in principle, but also many limitations in practice.

Current technology allows the execution of circuits with more than 100 qubits and a depth of about 60 two-qubit gates.\cite{kim2023evidence} While error correction is not yet available, there are error mitigation methods that enable unbiased estimation of the expectation value of operators.\cite{cai2022quantum,van2023probabilistic}
It seems, therefore, that all the ingredients for enabling NISQ variational methods  are present, as such a circuit can define a variational ansatz that could outperform the best classical ansatz for a given problem.
A central point of this Perspective is thus what is missing to translate this potential into practical variational computation and, hopefully, achieve quantum advantage.

The first point to establish is what kind of circuit can be used to create a ground state of our target many-body quantum Hamiltonian. As we have seen, the advantage of a quantum computer is the mere possibility of storing an exponentially large wavefunction with a linear number of qubits in memory (cfn. Eq.~\ref{eq:ket}). But this gives us no guarantee that (1) a quantum circuit with a finite, and possibly small, depth can give us a better approximation than the best classical method, and (2) even more importantly from a conceptual point of view, that it is possible to optimize the parameters even assuming that the ground state is contained in the variational ansatz.

An obvious drawback of variational calculations in the NISQ regime is the presence of noise. A gate error of $0.1\%$ propagating through a circuit with a depth of 100, composed of 100 qubits, produces a state with fidelity on the order of $10^{-3}-10^{-4}$ compared to the noiseless case.\cite{kim2023evidence} However, there exist error mitigation methods that can be applied to obtain unbiased estimates of expectation values.
Assuming, as a working hypothesis, that the noiseless version of a quantum circuit can generate a variational ansatz that is better than the best classically available, or even the exact one, everything becomes a game of exponents, as error mitigation incurs an exponential cost in circuit repetitions.\cite{cai2022quantum,van2023probabilistic,o2022purification} It remains to be determined whether the exponent of this post-processing step is mild enough to guarantee reasonable runtimes for classically intractable molecules.

In this Perspective, we will not focus on hardware error mitigation, which is introduced in the recent Ref.~\cite{o2022purification} but on another complementary issue.
As we will see, a major problem with quantum variational methods is even more surprising: even assuming that we have prepared exactly the target state, computing its energy is -so far- an inefficient procedure. Although this inefficiency is not the same as in  complexity theory definition, where an algorithm is said to be inefficient if it is exponentially scaling, it is from a practical point of view. This is a completely new condition that does not happen, for example, in variational calculation using QMC.

\subsection{The variance problem}
\label{ss:variance}

The fundamental (and obvious) concept at the root of this is that a quantum computer obeys the postulates of quantum mechanics: we cannot access the state that is created by the circuit exactly, but only through measurements. We can measure the expectation value of the Hamiltonian by post-processing the read-out of each measurement, and then prepare exactly the same state and repeat the measurement, and so on.

Let us suppose to have prepared a quantum state $\ket \psi$:
following  Eq.~\ref{eq:pauli_ham}, we see that the mean value of the Hamiltonian is the linear combination of the expectation values of the $N_p$ Pauli strings,
\begin{equation}
\label{eq:paulihamexpval}
    \bra \psi H \ket \psi := \langle  H \rangle = \sum_{j=1}^{N_P} h_j \langle P_j \rangle  
\end{equation}
For simplicity we assume that the expectation values of the $P_j$'s are obtained from $N_P$ independent sets of measurements: the error on the estimate is then given by
\begin{equation}
    \epsilon = \sqrt{ \sum_j |h_j|^2 \textrm{Var}[P_j] / M_j   },
\end{equation}
where $\textrm{Var}[P_j] = \langle P_j^2 \rangle - \langle P_j \rangle^2 \le 1$ is evaluated using $M_j$ repeated measurements, or \emph{shots}.

Since we need to evaluate each $\langle P_j \rangle$ independently, their statistical fluctuations are not correlated, so one needs to reach chemical accuracy by resolving each expectation value with very high accuracy as it  gets multiplied by a possibly very large prefactor $h_j$.
In Wecker et. al.\cite{wecker2015progress}, it is estimated that $ M = \sum_j M_j =10^9$ measurements would be needed to reach $\epsilon \sim 10^{-3}$ Ha, for $H_2O$, and up to $10^{13}$ for $Fe_2 S_2$ decribed with 112 spin-orbitals and STO-3G basis.

This happens even in the case where we prepare the exact ground state, thus violating the \emph{zero-variance} property of the ground-state if the energy is calculated this way.
This issue is often called \emph{the variance problem}, and is one of the most overlooked issues in the VQE community, which seems to be more active in new circuit ansatze development.
However, several works aiming to mitigate this problem have been put forward. The simplest one consists in grouping all $P_j$'s that commute qubit-wise and that can be measured simultaneously.\cite{kandala_hardware-efficient_2017}
Other methods aim to find better grouping schemes, introducing general commutativity rules at the expense of longer measurement circuits\cite{gokhale2019minimizing, crawford2021efficient, yen2020measuring, huggins2021efficient,PhysRevX.10.031064}.

For the evaluation of two-body reduced density matrices for fermionic systems, it is possible to devise an asymptotically optimal measurement scheme\cite{PhysRevX.10.031064,cotler2020quantum}, with a number of unique measurements circuits scaling as $N_{\textrm{groups}}\sim \mathcal{O}(n^2)$. However, there is still the problem that all these correlators, estimated stochastically, need to be summed up to $\langle H \rangle$. Therefore, $N_{\textrm{groups}}$ is not a faithful representation of the number of total measurements to be performed to achieve chemical accuracy, since each partition needs to be measured $M_{\textrm{per~groups~to~$\epsilon$}}$ times, as for the sum to achieve a total error of $\epsilon$.

There are also methods based on a different philosophy, namely to systematically \emph{approximate} the electronic Hamiltonian (Eq.~\ref{eq:fermionic_ham}) to reduce the number of terms in the sum, using a low-rank representation of the Hamiltonian.
\cite{motta2021low,oumarou2022accelerating}
This technique finds application also to real-time dynamics simulations, and may considerably reduce the runtime of error-corrected algorithms\cite{goings2022reliably}. While it can certainly mitigate the variance problem in the NISQ era, it does not qualitatively solve the problem for the very same reason outlined above.
We observe that several works aim to reduce the number of basis states required to represent the Hamiltonian. However, this may not necessarily improve the number of measurements, as these fewer terms may have a larger variance.
An interesting example of this is seen in variational quantum algorithms applied to classical cost functions. This scenario is important for optimization, and in this case, the most popular variational method is called the Quantum Approximate Optimization Algorithm (QAOA). Despite the fact that the cost function, by definition, only needs to be measured in one basis—the computational basis—the impact of shot noise is still quite detrimental to the overall performance even in this case.\cite{scriva2023challenges}

Finally, we also report the celebrated \emph{shadow tomography}\cite{huang2020predicting} method, which may be useful for estimating local qubit operators but has an exponential scaling for non-local ones, such as our $P_j$'s.

In general, the total number of circuit repetitions during an optimization run based on energy optimization, featuring $N_{\textrm{iter}}$ optimization steps is
\begin{equation}
    M_{\textrm{VQE}} = N_{\textrm{groups}} \times M_{\textrm{per~groups~to~$\epsilon$}} \times N_{\textrm{iter}}.
\end{equation}

Let us consider a concrete example of a $H_2O$ molecule, in a very minimal basis consisting of 12 spin-orbitals. Following a state-of-art variance reduction method\cite{huggins2021efficient}, we quote a number of $10^8$ circuit repetitions to compute a single point energy within $10^{-3}$ Ha accuracy (notice that this is the error from the exact value obtained with this minimal basis set, not the exact value using a converged basis set).
Assuming a circuit execution time, with measurements, of $1 \mu s$, we are limited to about hundreds of optimization steps per day (totally neglecting classical communications and reset times).

Notice that NISQ hardware means that hardware noise will always be present, and this noise usually varies with time, that's why an optimization run that lasts for more than one day will likely never converge: the optimal parameters $\bm{\theta}$ which yields the lowest energy with the hardware configuration of today, may not be optimal the day after.
The fact that hardware errors are device and time-dependent realistically excludes the possibility to parallelize the shots using several machines, as is customary in conventional QMC. In this case, one would claim the possibility  to prepare always the same trial state on different noisy machines.

\subsection{The noisy optimization problem}
\label{ss:optimization}

The second step of any numerical variational calculation is optimization. In this perspective, we aim to maintain a high-level tone and will not be concerned with details such as which optimization method is better or worse depending on the type of circuit. In general, optimizing the parameters is a complicated task that is delegated to the classical part of the algorithm.
Heuristic circuits are very short but have a much larger number of variational parameters than those inspired by chemistry, such as the unitary coupled cluster, or physics, such as the Hamiltonian variational ansatz.\cite{Cao_2019} The latter have longer circuits but allow for a more stable optimization.

A wealth of literature focuses on theoretical roadblocks, such as the existence of barren plateaus\cite{barren2018} and the fact that optimization itself is an NP-hard problem.\cite{bittel2021training}
However, we observe that even in the conventional case, the optimization of parameters occurs in a corrugated landscape. Nevertheless, it is almost routine to optimize thousands of variational parameters in variational Monte Carlo (VMC).\cite{nakano2020turborvb} Moreover, barren plateaus are a concept borrowed from the quantum machine learning community and is likely not relevant, or at least not the real bottleneck, in the case of using structured ansatzes\cite{bosse2022probing,astrakhantsev2022algorithmic,martin2023barren} (i.e. non-random or heuristic), which should be the norm for studying physical systems.
The reason why this concept has never arisen in conventional VMC is that no one has ever tried to optimize molecular systems from quasi-random trial states.

In this Perspective, we remain faithful to our practical approach and briefly analyze the problems arising from the simple existence of statistical noise.
First, we observe that since the zero variance property does not hold, we can only use the energy and not its variance as a cost function. Optimizing using finite differences is inefficient since each step is affected by statistical noise. Typically, in VMC, this problem can be solved through correlated sampling\cite{becca_sorella_2017}, which is not possible in this case.
The other possibility is to calculate the expectation value of the generalized forces $f_i$, defined as the derivative of the energy with respect to the variational parameters. For simple circuits, calculating the force is possible thanks to a technique called the parameter shift rule\cite{mcardle2018quantum}, and extensions are possible for more structured circuits.\cite{astrakhantsev2022algorithmic} Due to the no-free-lunch theorem, the statistical error that we had using energy translates into statistical error on the forces.
The optimization effectively becomes a stochastic gradient descent, which resembles a discrete Langevin equation at finite effective temperature,
\begin{equation}
\theta_i^{'}= \theta_i + \delta f_i + \eta_i^{\textrm{shot}},
\end{equation}
where $ f_i = -\partial_i \langle  H \rangle$, $\eta_i^{\textrm{shot}}$ is a Gaussian distributed random variable, and $\delta$ is a finite integration step.
Astrakhantsev et. al.\cite{astrakhantsev2022algorithmic} have shown that the statistical error defines an effective temperature, $T^{\textrm{shot}}$, proportional to the variance of the random variable $\eta_i^{\textrm{shot}}$.
Below a certain number of samples $M^*$, therefore above a certain effective noise temperature the search is unsuccesful. Above, the optimization becomes possible.
Moreover in the $M \gg M^*$ regime, the infidelity of the state preparation seems to scale as $1/\Delta^2$, where $\Delta$ is the energy gap between the ground and the first excited state.
These numerical results have been obtained on a challenging $j_1-j_2$ Heisemberg models (cfn. Ref.~\cite{wu2023variational}), and it would be interesting to check how they generalize to chemistry problems.
On the optimistic side, the critical number of samples $M^*$ seems to not scale exponentially with the system's size, though allowing in principle efficient VQE optimization in the presence of quantum measurement noise.
Moreover, in this state-of-the-art VQE study, barren plateaus are not observed.
\\

Another point of contact between variational quantum computing and conventional Monte Carlo is that techniques well-known in the latter for decades are slowly being adopted in this new field. Perhaps one of the most important is the use of the so-called quantum information matrix to precondition the gradient at each step with the following matrix,
\begin{equation}
\label{eq:S}
    S_{ij} = \braket{ \partial_i \psi | \partial_j \psi } -  \braket{ \partial_i \psi |  \psi }  \braket{  \psi | \partial_j \psi },
\end{equation}
where $\ket{\partial_j \psi }$ is the derivative of $\ket \psi$ as a function of the $j$-th variational parameter.
Although most of the community believes that Eq.~\ref{eq:S}  comes from machine learning, where it is used in the \emph{natural gradient} ,\cite{Amari1998} it has actually been used for more than twenty years in VMC to optimize trial wave functions for chemistry and condensed matter.\cite{becca_sorella_2017} It was introduced by Sorella in the stochastic reconfiguration method\cite{sorella1998green,sorella2001} and later given the geometric meaning of metric of the space of variational parameters.\cite{2012_mazolla} A weak regularization of the diagonal is sufficient to obtain stable and effective optimizations, as has also been shown in the quantum case.\cite{stokes_quantum_2020}

However, the measurement problem also heavily affects the calculation of algorithm efficiency in this case. In VMC, the matrix $S$ can be evaluated with negligible overhead using the same samples $x$'s, distributed as $|\psi(x)|^2$ (which can be evaluated classically there), that were already generated for the energy calculation. In the quantum case, each matrix element must be statistically evaluated using (uncorrelated) repetitions of a specific circuit for the pair $i,j$. Moreover, it is not trivial to obtain a circuit for each element $S_{ij}$, and only a block-diagonal approximation of $S$, where $i$ and $j$ belong to the same block, is the most feasible solution.\cite{stokes_quantum_2020}

At the moment, an interesting development to overcome this problem is a heuristic combination of stochastic reconfiguration with the SPSA optimizer. In this case, the matrix S, which would require $N_{par}^2$ circuits, is approximated by a Hessian calculated using only two random directions in the parameter space.\cite{gacon2021simultaneous} However, the numerical benchmarks proposed to validate the method are too small to fully understand the real possibilities of this simplified optimizer.

Finally, another connection between VQE and QMC arises in the context of using noisy ionic forces in molecular dynamics (MD) simulations. In such cases, it becomes impractical to follow an energy-conserving trajectory. Sokolov et al.\cite{sokolov2021microcanonical} utilize a similar technique proposed in QMC-powered MD simulations from many years ago.\cite{attaccalite2008stable} They use shot noise to define an effective Langevin MD, enabling unbiased simulations at constant temperature.

\section{Fault-tolerant quantum chemistry}
\label{s:fault_tole}

At this point, we would face the conundrum that a quantum computer can in principle store an exact wavefunction, but we cannot practically evaluate its energy or other expectation values using the variational methods introduced so far. However, quantum computing admits an efficient method for calculating energy, even more efficient than Monte Carlo methods. These methods are based on the quantum phase estimation (QPE) algorithm or its successive variants.

The standard version of this algorithm,\cite{nielsen_chuang_2010} which finds applications far beyond chemistry, works as follows.
Suppose we have a unitary operator $U$ and one of its eigenstates $\ket{\psi_n}$. We can evaluate its eigenvalue $\lambda_n = e^{i 2\pi \phi_n}$ (expressible as a function of its phase since $U$ is unitary) using an extra register of $r$ qubits and $r$ controlled operations $\textsf{c}U$, where the first operation  controls the application of $U$, the second of $U^2$, the third $U^4$, and so on, until the final $U^{2^{r-1}}$.  Finally, it is sufficient to apply a quantum Fourier transform to read the phase value in binary form, truncated to $r$ bits.

To arrive at an implementation that interests us, we simply identify this generic unitary operator $U$ with $e^{iHt}$, i.e., a Hamiltonian evolution operator, and the starting state as an eigenstate of $H$, for example, the ground state. In this case, the phase we read is $E_n t$.
The operation $U^{2^{r-1}}$ translates into an evolution of time $t~2^{r-1}$.

It is possible to show that the error (due to truncation in $r$ bits) we make on the phase $E_n t$ scales as $1/r$, where $r$ is the total number of applications of $U$.
This is because the discretization error scales as $2^{-r}$, but each application of the controlled unitary doubles the length of the circuit.
In literature, this is often quoted as a quadratic speed-up compared to a Monte Carlo evaluation, whose error scales as $1/\sqrt{M}$, where $M$ is the number of samples, and thus the number of function calls of the function to be evaluated on the generated distribution. If we interpret $r$ as the number of ``function calls'' i.e., $\textsf{c}U$ operations applied to the state $\ket{\psi_n}$, the asymptotic comparison with Monte Carlo can be made, keeping these specifications in mind.

There are now two complications to consider. The controlled operation $e^{iHt}$ cannot be implemented exactly but requires approximations. The most established is the Trotter step decomposition, which has the advantage of not requiring additional qubits.\cite{lloyd:1996,childs:2021} Recently, other methods that have better asymptotic scaling but require additional qubits, such as the linear combination of unitaries\cite{childs2012hamiltonian} and qubitization\cite{low2019hamiltonian}, have surpassed Trotter's method in popularity. 
For example, the first runtime and resource estimation works for chemistry, by Troyer and coworkers\cite{Reiher_2017}, assumed that QPE with Trotter had to be used, while now more recent works use qubitization (plus many other tricks to shorten the circuit).\cite{goings2022reliably}
Notice however that the empirical performance of Trotter methods can be better than predicted upper bounds, and this is still an active  area of research.\cite{childs2021theory,ostmeyer2022optimised,zhuk2023trotter}

The first crucial observation is that now even a ground-state calculation requires a black-box that implements real-time dynamics, or closely matching objects. This brings us back to the initial discussions about the complexity of implementing Hamiltonian simulations of fermionic systems, much more complex than their spin-lattice counterparts.
Given that we require quantitative precision on the time evolution, the Hamiltonian evolution 
algorithm requires a full fault-tolerant implementation.

The second observation is that these algorithms require the challenging assumption of having the ground state as their input. What happens if the state on which we apply QPE estimation is not the ground state? Here there is good news and bad news. Let's start with the good news: unlike the classical case if we input a generic state $\Phi$ (classically, the equivalent would be preparing a generic ansatz and sampling with Metropolis from it), when we read the phase register, the state must collapse onto an eigenstate of $\ket{\psi_n}$ and the read-out phase is $\phi_n$. Therefore, the energy readout in the auxiliary register determines the collapse onto an eigenstate of $H$ of the previously initialized state $\Phi$. The good news is therefore that we will not read a random number, but one of the possible eigenvalues. The bad news is that we do not know which one. In general, we will read the eigenvalue $n$ with a probability given by the overlap $|\braket{ \Phi | \psi_n }|^2$. It then becomes crucial that, if we are interested in the ground state, the initial state is not completely random, but has a sizable overlap with the ground state.

Generally, in chemistry and materials science, we are interested in the runtime scaling with size, i.e., number of electrons, basis set, etc. If the overlap vanishes exponentially with size, the entire procedure becomes exponentially long, nullifying the exponential advantage that could be initially envisioned, given the compression in memory and the possibility of efficiently reading the energy. A recent study focused on this aspect, showing that this issue, though known in principle but often forgotten in practice, could seriously undermine the claim of exponential advantage for electronic structure.\cite{lee2023advantagechemistry}
It should be noted, however, that even a polynomial advantage could be sufficient to solve problems that are still intractable, as can be seen from how Density Functional Theory has revolutionized chemistry and materials science, thanks to its improved $N^3$ scaling compared to the $N^6,N^7$ scalings of coupled-cluster.

To conclude, the absence of an exponential speed-up does not rule out the existence of a practical quantum advantage, which is more difficult to identify a priori, but on a case-by-case basis. In this context, resource estimation studies focusing on particular molecular systems are of great importance.  
Goings et. al.~\cite{goings2022reliably} perform resource estimates to simulate a challenging system for classical methods, the cytochrome P450 enzyme.
The estimates depend on the hardware noise that needs to be corrected. To simulate the ground state using an accurate active space, $\sim 5 \times 10^6$ ($5 \times 10^5$) physical qubits with error rates of $0.1~\%$ ($0.001~\%$) would be needed.
Concerning materials science systems, state-of-the-art studies are represented by Rubin. et. al.~\cite{rubin2023fault}, and Ivanov et. al.~\cite{PhysRevResearch.5.013200}, which move away from the plane-wave basis set and combine Bloch, or Wannier orbitals, respectively, with most recent techniques such as sparse qubitization or tensor hyper-contraction.
Resource estimates applied to  Lithium Nickel Oxide battery cathode\cite{rubin2023fault}, and transition metal oxides\cite{PhysRevResearch.5.013200} indicate longer fault-tolerant runtimes compared to molecular systems such as P450.
Clearly, such estimates are based on state-of-art algorithms, including the most efficient way to encode fermionic Hamiltonians for phase estimation, and the current state of error correction algorithm.
Further algorithmic developments will improve the cost of the simulations~\cite{vonBurg2021catalysis}.
Orders of magnitude in efficiency have been gained compared to just ten years ago,\cite{whitfield2010simulation} and therefore the threshold for quantum advantage could shift in one direction or another, approaching when new quantum strategies are invented or perhaps moving away thanks to the constant progress of ``conventional'' methods such as DMRG or QMC.

\section{The local energy in variational Monte Carlo}
\label{sect:vmc}

After extensively introducing quantum computing algorithms for chemistry and specifically discussing the practical limitations of variational approaches, let's move on to the classical case.
It's very instructive to understand why classical variational methods do not suffer from the same variance problem, to guide us in inventing equally efficient energy estimators.
Let's start again from the formal definition of the energy's expectation value over a general (unnormalized) state $\ket \psi$
\begin{equation}
\label{eq:expvalint}
   E=  \langle  H \rangle = { \bra \psi H \ket \psi  \over \braket{\psi|\psi} } = {  \sum_x \braket{\psi | x}   \bra x H \ket \psi   \over  \sum_x |\braket{\psi | x}|^2  },
\end{equation}
where we insert $\sum_x \ket x \bra x$ in the denominator and numerator.
Notice that here we use the notation for a discrete Hilbert space, but the formula can be generalized to continuous models by replacing the sum with the integral ($\sum_x \rightarrow \int dx$).
Some steps are necessary to transform the Eq.~\ref{eq:expvalint} into the typical Monte Carlo format, where we integrate the product of a probability distribution from which we can sample form, $p(x)$, and an objective function.
This is achieved formally by dividing and multiplying by $\psi(x) =\braket{x|\psi}$.
Eq.~\ref{eq:expvalint} then becomes
\begin{equation}
E =    {  \sum_x |\braket{\psi | x}|^2   E_L (x)   \over  \sum_x |\braket{\psi | x}|^2  } = \sum_x p(x) E_L(x),
\end{equation}
where the \emph{local energy}
is defined as
\begin{equation}
    E_L(x) = { \bra x H \ket \psi   \over \braket{x|\psi}  },
\end{equation}
and is a quantity that can be evaluated locally for the configuration $x$ (see below).
The probability distribution is defined as $ p(x) =  {  |\braket{\psi | x}|^2    / \sum_x |\braket{\psi | x}|^2  }  $.
Now, if we assume that we can sample configurations $x \sim p(x)$, i.e. using a Markov-chain algorithm, the energy can be evaluated stochastically as
\begin{equation}
\label{eq:statsum}
    E \approx {1 \over M_{\mathrm{VMC}} } \sum_{i=1}^{M_{\mathrm{VMC}}} E_L(x_i),
\end{equation}
using $M_{\mathrm{VMC}}$ decorrelated samples taken from a Markov-chain algoritm, such as Metropolis\cite{metropolis1953equation}.
This technique is called Variational Monte Carlo (VMC)\cite{becca_sorella_2017} and is efficient as long as
\emph{(i)} computing $E_L(x)$ is efficient, and \emph{(ii)} it is possible to run the Metropolis algorithm also efficiently. Since classical trial functions $\psi(x)$ can be evaluated with numerical precision, for each $x$, then it is also the ratio $ | \psi(x') |^2 / | \psi(x) |^2 $ for each pair $x,x'$, which is needed to perform a Metropolis update.\footnote{assuming a symmetric proposal step, then one just need the ratio of probabilities to update the walker $p(x')/p(x)$ and perform the acceptance test.}

One of the most important features of the local energy is that its variance is zero when $\ket \psi$ is an eigenstate of $H$.
Indeed, if $H \ket {\psi_0} = E_0 \ket {\psi_0} $, then 
$E_L (x) =  \bra x E_0 \ket \psi_0 / \braket{ x| \psi_0} = E_0$.
In practice, this means that the local energy function will be closer to a constant value as the trial state $\ket \psi$ approaches the ground state. This results in reduced statistical fluctuation in Eq.~\ref{eq:statsum}.

At the same time, it is possible to use the variance of the local energy as a cost function for the optimization. This allows in principle to certify the success of the minimization, as the ground state is signaled by zero statistical error.

\subsection{The local energy in practice}

The calculation of the local energy depends on the model and the wave function.
In continuous space, the wave function can be given by a Slater determinant ansatz (for fermionic systems), usually complemented with an explicit correlation operator like the Jastrow factor.\cite{McMillan1965,becca_sorella_2017}
In this case, evaluating the local energy reduces to applying the Laplacian operator to the function in real space and dividing by the function itself.
For the sake of clarity, let's consider a toy example. The local energy for a (unnormalized) Gaussian trial ansatz $\psi(x)= e^{-\theta x^2} $, in continuous space, and a typical one-dimensional Hamiltonian
$ H = -1/2 (\partial^2 / \partial x^2) + V(x)$,
reads
\begin{equation}
    E_L(x) = \theta - 2 \theta^2 x^2 + V(x).
\end{equation}

The local energy depends on $x$ and the variational parameter $\theta$, which can be optimized in an outer loop. In this case, it can be observed that if the external potential is a harmonic oscillator $V(x) = \omega/2 ~x^2$, the local energy becomes 
\begin{equation}
    E_L^{\textrm{h.o.}}(x) = \theta + x^2 \left( \frac{\omega}{2} - 2\theta^2 \right).
\end{equation}
The local energy no longer depends on $x$ when the variational parameter takes the value $\theta = \omega/2$, for which the variational ansatz becomes exact.
Moreover, it also takes the value $ E_L^{\textrm{h.o./opt.}} = \omega/2$ with zero statistical fluctuations in Eq.~\ref{eq:statsum} as $E_L$ does not depend on the sampled point $x_i$ anymore.
Modern codes for solving the many-electron Schrödinger equation in chemistry or materials science feature sophisticated trial ansatz, which are in turn functions of atomic orbitals.\cite{nakano2020turborvb}
While in the past, introducing a new ansatz required coding new functions for the evaluating the derivatives, now, the evaluation of the local energy can be delegated to algorithmic differentiation routines. This allows for the adoption of fairly sophisticated ansatze in VMC.\cite{sorella2010algorithmic,nakano2020turborvb}

The local energy shows up every in VMC calculation including lattice models such as spin and Hubbard models.
In this case, the spatial derivatives are replaced by non-diagonal quantum operators such as spin-flip or hopping operators, $H_{x,x'} = \langle x' | H | x \rangle$, 
where $x,x'$ can represent a specific spin configuration or an occupation state of fermions or bosons on a lattice.
In this discrete basis, the local energy is written as follows,
\begin{equation}
\label{eq:le_discrete}
    E_L(x) = { \langle x | H | \psi \rangle \over  \langle x  | \psi \rangle } =  { \sum_{x'} H_{x',x} \langle x' |  \psi \rangle \over  \langle x  | \psi \rangle }. 
\end{equation}
and can be computed efficiently as long as the number of states $x'$ such that the Hamiltonian matrix elements $|H_{x,x'}| \neq 0$, at fixed $x$,  is only polynomially increasing.

\subsection{Pauli measurements versus local energy}
\label{ss:numerics}

To better illustrate these concepts, it can be instructive to perform a numerical experiment on a toy model, the one-dimensional transverse field Ising model,
\begin{equation}
\label{eq:tfim}
    H = H_1 + H_2 = - J \sum_{k=1}^L \sigma^z_k \sigma^z_{k+1} - \Gamma \sum_{k=1}^L \sigma^x_k ,
\end{equation}
where $\sigma^\alpha$ are Pauli matrices,
and consider the critical transition point at $J = \Gamma =1$. We also consider a short chain of $L=10$.
We denote a generic computational basis configuration as $|x\rangle = (s_1,\cdots, s_L)$, where $s_k$ are eigenvalues $\{1,-1\}$ of the $\sigma_j^z$ operator. 

In this case, the spin Hamiltonian is already expressed in Pauli terms (one just needs to re-define the eigenvalue of the spin-$z$ operator from $\{1,-1\}$ to $\{0,1\}$).  Regarding the VQE approach, the energy can be measured in only two bases: the computational basis and the ``$XX\cdots X$'' basis, obtained by applying a Hadamard gate, \textsf{H}, on each qubit at the end of the circuit that prepares the variational state.

In this numerical experiment, we use the variational Hamiltonian form, with a sufficiently deep circuit of up to 24 layers, resulting in up to 48 variational parameters (see Appendix~\ref{app:details}). By optimizing ansatze characterized by different circuit depths (without shot noise for simplicity), it is possible to obtain trial states systematically closer to the exact ground state of the model.\cite{wu2023variational} In Fig.~\ref{fig:local_energy}, we use depths ranging from 12 to 24, and we can reach a relative error on the energy of $10^{-5}$ compared to the exact ground state energy, $E_0$.

However, the statistical error on the energy, which is evaluated with Eq.~\ref{eq:paulihamexpval}, does not improve. In fact, if we had tried to optimize the circuit using the noisy energy estimator, we would not have been able to obtain such accurate optimized trial states. This clearly demonstrates that the estimator does not possess the zero variance property, as opposed to the VMC calculation.

To obtain the standard deviation in Fig.~\ref{fig:local_energy} we repeat the estimation of the variational energy, Eq.~\ref{eq:paulihamexpval} (Eq.~\ref{eq:statsum} for the VMC case described below), 100 times to obtain a population of variational energies that could be obtained with the given variational setting, $M_j$ ($M_\textrm{VMC}$ for the VMC case) setups.
We use a number of shots $M_j, M_\textrm{VMC}$, which is smaller ($10^2$), equal ($10^3$), and larger ($10^5$) than the Hilbert space of the model, i.e. $2^{10} =1024$.

For the VMC comparison, we deliberately use a fairly simple classical ansatz, a long-range Jastrow state, which features only 5 variational parameters for $L=10$ (see Appendix~\ref{app:details}). Although this classical ansatz only reaches a moderate relative accuracy of $10^{-3}$, at best, the statistical error on the energy consistently improves, outperforming the statistical error obtained with the quantum circuit.
Notice that this is an easy model for VQE: the number of measurement basis is the minimum possible for a genuine quantum many-body system.
Electronic structure Hamiltonians unfold into thousands of Pauli operators, which in turn require similar numbers of basis.
This numerical example demonstrates the power of the local energy-based estimator compared to the Pauli measurement one.
From this example we can understand also the following lesson: even finding the smallest set possible of basis to measure $H$ will not solve all our problems, as this estimator still lacks the zero-variance property.

A hybrid solution has been proposed by Torlai et. al.~\cite{torlai2020precise}. They use quantum state tomography, using neural-networks\cite{torlai2018neural}, and a tomographically incomplete basis set, to obtain a classical reconstruction of the quantum state. Classical VMC can be then applied to this classical approximation to calculate, precisely, the energy.
This method solves the variance problem but it introduces a bias stemming from a possibly, and likely, imperfect reconstruction of the quantum state.
Moreover, it raises the question of finding the range of applicability of the method. If the quantum states can indeed be represented by a classical ansatz, then one could directly reach the ground state by optimizing that, without the need of a quantum computer.

\begin{figure}[ht!]
\includegraphics[width=\columnwidth]{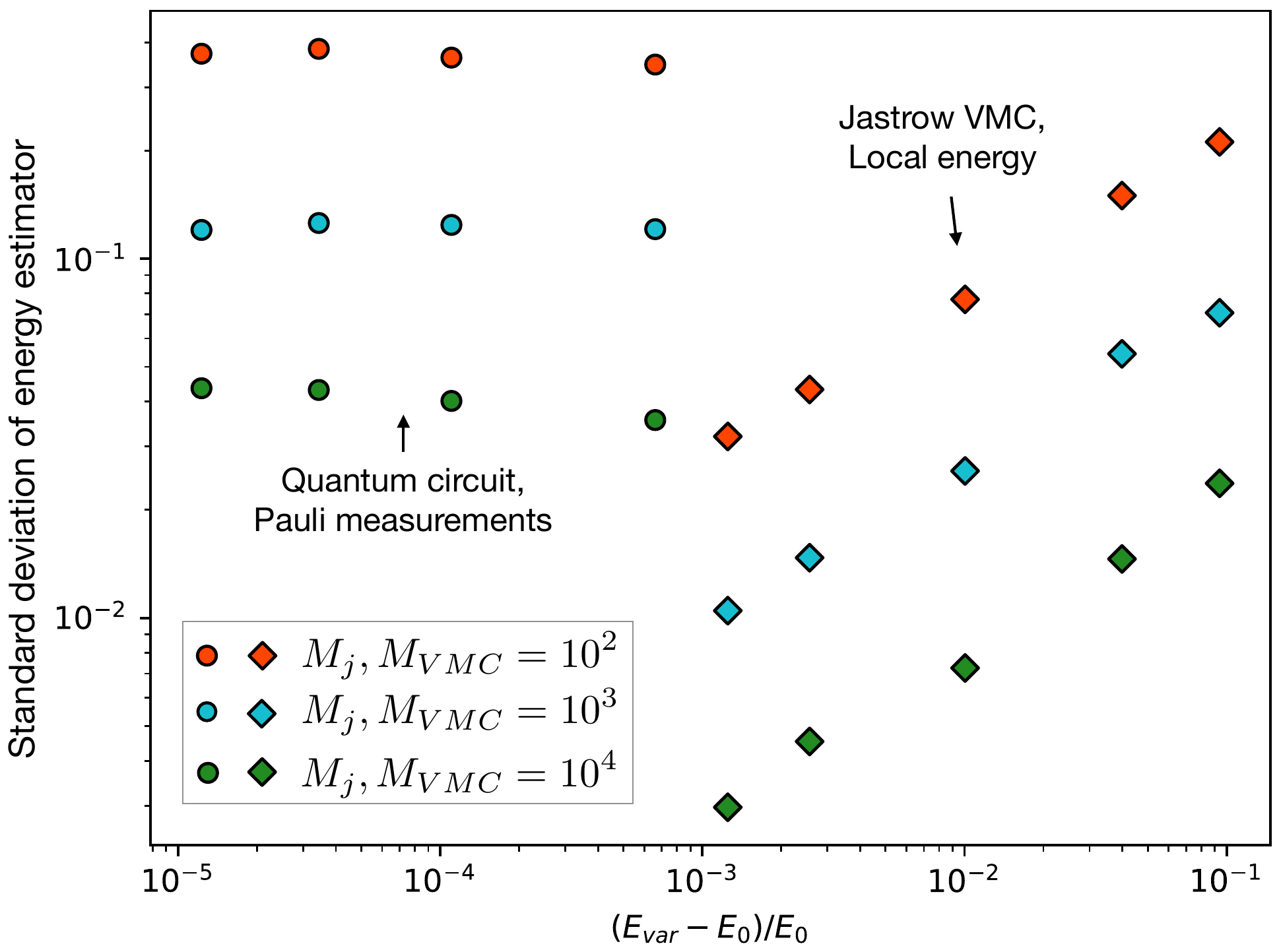}
\caption{ Standard deviation of the energy estimator using a quantum circuit and Pauli measurements, Eq.~\ref{eq:paulihamexpval} (circles), and with a simple classical ansatz but using the local energy, Eq.~\ref{eq:le_discrete} and Eq.~\ref{eq:statsum}(diamonds). Different colors indicate different sampling sizes. In the quantum case, the dataset is made of $M_j$ wavefunction collapses per basis (which are two for the model considered), while for the classical case, it is made of  $M_{\textrm{VMC}}$ spin configurations, $x_i$, sampled from the trial state $|\psi(x_i)|^2$.
In both cases, we prepare different ansatze, within the same ansatz class, but having different accuracies.
For each trial state, we plot the standard deviation of the energy vs. the relative error of its variational energy $E_{\textrm{var}}$ (computed exactly).
The zero variance property only holds in the VMC case, since the statistical error of the quantum energy estimator remains finite even when the trial state approaches the exact limit.
}
\label{fig:local_energy}
\end{figure}

\subsection{Beyond variational: projective methods}
The local energy is a central concept in quantum Monte Carlo beyond the simplest VMC method because, in practice, every projective QMC method requires a trial wave function $\psi$ to alleviate the sign problem in fermionic simulations or, more generally, to reduce statistical fluctuations. These projective methods, such as Diffusion Monte Carlo\cite{becca_sorella_2017} or Auxiliary Field Monte Carlo (AFQMC)\cite{motta2018ab}, improve upon the VMC energy but still rely on a variational state for importance sampling. Thus, the local energy resurfaces in these contexts as well. An accurate QMC simulation is rarely seen without a good variational starting point.\footnote{S. Sorella, private communication.}

One can therefore see a similarity between the importance of VMC, which is foundational for a more accurate calculation with projective QMC, and the significance of the initial state preparation for a successful execution of QPE in quantum computing. It is highly likely that this duality between variational and projective methods (in imaginary time in the classical case and in real-time in the quantum case) will extend to quantum computing. In that case, algorithms like VQE or its variants, despite being considered already old-fashioned by some, will remain central even in the fault-tolerant regime as state preparation subroutines.

\section{Quantum computing meets quantum Monte Carlo}
\label{sect:qcqmc}

\subsection{The local energy in quantum computing}

The existence of a local energy estimator in quantum computing would eliminate any variance problem in VQE. However, it is not as straightforward to apply this trick in quantum computing simply because evaluating the ratio $ \braket{x'|\psi} / \braket{x|\psi}$ becomes extremely demanding in general.\footnote{here
we consider the discrete basis definition of the local energy, Eq.~\ref{eq:le_discrete} since it is more appropriate for the digital qubit space. Here $\ket x$ can be a basis vector $\ket i$, defined in Eq.~\ref{eq:ket}. }

Here $\psi(x) = \langle x| \psi \rangle $ needs to be evaluated from quantum measurements, hence is affected by statistical noise. 
While evaluating  $\psi(x)$ to additive precision is possible, the local energy involves ratios of amplitudes. Maintaining a fixed precision on the ratio 
\begin{equation}
\label{eq:ratio}
    \braket{x'|\psi} / \braket{x|\psi}
\end{equation}
 is costly, because quantum states have generally an exponentially large support. This translates into exponentially vanishing amplitudes at the denominator of Eq.~\ref{eq:le_discrete}.

These statistical fluctuations are different compared to those found in a standard Monte Carlo calculation. In VMC, the local energy can always be computed with numerical precision and the fluctuations arise from a finite number of samples, $M_\textrm{VMC}$ in Eq.~\ref{eq:statsum} (in the presence of an approximate trial state). Here uncontrolled statistical fluctuations arise solely from the estimation of the local energy at a fixed $x_i$.

We are witnessing an increase in works that aim to combine quantum computing and quantum Monte Carlo.  Huggins and coworkers proposed an interesting combination of quantum computing and AFQMC.\cite{huggins2022unbiasing}
In this work, they use a circuit to generate the trial wave function, from which samples are drawn (in this representation, the configuration $x$ is a Slater determinant). The AFQMC algorithm then proceeds unchanged, and the supposed advantage of the method lies in using a circuit to generate an ansatz that could be inaccessible classically.
Mazzola and Carleo\cite{mazzola2022exponential} showed that the procedure, when adapted to many-body lattice models at criticality, thus using Green's function Monte Carlo instead of AFQMC, exhibits an exponentially scaling behavior with a hard exponent. This is due to Eq.~\ref{eq:ratio} and that strongly-correlated states have vanishing overlaps on the configuration's basis, necessitating an exponentially increasing number of samples to compute the local energy.
It is estimated that a reasonably accurate ground state calculation of a 40-sites transverse-field Ising model (Eq.~\ref{eq:tfim}) requires the order of $10^{13}$  measurements.
Inserting a gate frequency of 10 kHz (i.e. assuming a fault-tolerant implementation, see Sec.~\ref{s:hardware}) and a circuit depth of $\mathcal{O}(10)$ layers to generate an accurate trial state, this implies a runtime  of a few thousand years, for a system in reach of exact classical diagonalization.

Other works appeared  almost simultaneously
last year on this topic.
Zhang et. al~\cite{zhang2022quantum} introduced a quantum computing adaptation of FCIQMC\cite{booth2009fermion}. In this work a quantum circuit $U$ is used to create a `quantum' walker $\ket{\tilde{x}} = U \ket {x}$, i.e. a linear combination of Slater determinants $\ket{x}$, undergoing the FCIQMC subroutine.
The idea is interesting as it could counter the exponential explosion of the determinants/walkers during the imaginary time projection, by compressing logaritmically the memory to store them.
A possible major drawback of this method is that the Hamiltonian $H_{\tilde{x},\tilde{x}'}$ in this new basis is not sparse anymore.
Xu and Li\cite{xu2022quantum}
proposed to use Bayesian inference to reduce the number of shots required
to compute the local energy.
Kanno et. al.\cite{kanno2023quantum} further combines the ideas of Ref.~\cite{huggins2022unbiasing} with tensor networks.
Yang et. al.~\cite{PRXQuantum.2.040361} propose a way to speed-up \emph{real}-time path-integral MC already on NISQ hardware.
Tan et. al.\cite{tan2022sign} devise instead the integration with the Stochastic Series Expansion, another flavour of QMC used for spin models.

Finally, two recent works propose to use quantum data in a conventional VMC framework. In this case the local energy is calculated in conventional hardware.
Montanaro and Stanisic~\cite{montanaro2023accelerating} propose the usage of a VQE circuit as importance sampler to speed-up the first iteration of a VMC simulation.
Moss et. al.~\cite{moss2023enhancing} use quantum data from Rydberg atom simulators to train a classical neural-network ansatz (as in Ref.~\cite{torlai2020precise}) and further optimize it in a VMC fashion.

Overall, it is likely that an efficient way to estimate the local energy is possible only for sparse states, i.e for which the number of non-zero overlaps $\braket{x|\psi} \neq 0$ grows only polynomially with the system's size. However, it remains to be understood if a quantum computer is really needed to tackle such systems at this point.~\cite{schwarz_simulating_2013}
Furthermore, if a suitable basis transformation $U$ can be found to reduce the support of such states, then (1) this transformation should not spoil the sparsity of the Hamiltonian $H_{x,x'}$ to keep the evaluation of Eq.~\ref{eq:le_discrete} efficient. (2) Moreover, if this transformation exists it can be used to diagonalize efficiently the systems in a reduced sub-space without the need of QMC.

On a more positive note, it is not excluded that, despite exhibiting exponential scaling, the aforementioned approaches could yield a better exponent than the best classical method for some specific fermionic systems. To achieve this, it will be crucial to start with a classically intractable trial state to justify the subsequent imaginary-time projection. Further research and methodological advancements are required to assess the true potential of the method, in the presence of shot noise.

Overall, the pursuit of an efficient method for calculating energy, inspired by the local energy in VMC, is a field of research that we hope will yield numerous fruitful results. It is necessary for the quantum computing and QMC communities to clearly understand the limitations and potentialities of their respective techniques in order to invent new hybrid algorithms at the interface of these two worlds.

\subsection{Classical-inspired circuits for VQE, quantum-inspired ansatze for VMC}

The techniques and methods that have been used for decades in QMC are so numerous that many have been (and many are waiting to be) exported to quantum computing. Trial functions play a central role in VMC.
The use of explicitly correlated non-separable ansatze has brought great success to VMC and is basically a clever solution to compress the electronic wavefunction, which, when described in the space of determinants, requires otherwise an exponential number of coefficients. The latest iteration of this concept is the introduction of neural network quantum states by Carleo and Troyer in 2017,\cite{carleo_solving_2017} which can be seen as more general forms of Jastrow,\cite{McMillan1965,torlai2018neural} back-flow,\cite{backflow1981,Luo2019backflow} and tensor network states.\cite{gao2017efficient}

As mentioned earlier, compressing the Hilbert space within a polynomial scaling size quantum memory enables the manipulation of linear combinations of arbitrarily large Slater determinants. However, when considering the variational approach, we are constantly seeking shorter quantum circuits that can capture as much entanglement as possible, within the coherence time limitation of NISQ systems.

Several works have already proposed ways to implement a Gutzwiller operator, which is essentially the simplest form of a Jastrow operator, as a quantum circuit. Murta and Fernandes-Rossier\cite{murta2021gutzwiller} propose a method based on post-selection. Typically, the way to create non-unitary operators in quantum computing is through embedding them in a larger system that undergoes unitary evolution, a method also known as ``block encoding''. This involves introducing ancillary qubits, and it can be certified that the non-unitary operator has been successfully applied to the quantum state if and only if the ancillary register is measured and read in a given state. However, the problem with this approach is that the success probability decreases with the system size, requiring many repetitions.

Seki and coworkers\cite{seki2022gutzwiller} also propose a similar approach, based on the linear combinations of unitaries, and therefore also affected by a finite success probability.

Using a different approach, Mazzola and coworkers\cite{mazzola2019nonunitary} defined implicitely a hybrid quantum-classical wavefunction with a Jastrow operator in post-processing. The approach has been then improved in its scalability in Ref.~\cite{zhang2022variational}. There, a quantum circuit is used as importance sampler, and the measured configurations undergo post-processing by a neural-network.
Benfenati and coworkers\cite{benfenati2021improved} instead implemented a Jastrow operator moving it from the wavefunction to the Hamiltonian. This approach also do not require additional circuits compared to a VQE calculation. However, the re-defined Hamiltonian operator features much more Pauli terms to measure.
Motta and coworkers devised an imaginary time evolution (QITE) operator without ancilla and post-selections.\cite{motta2020determining}
The original formulation of the method formally incurs an exponential dependence on the correlation length in the general case, because it requires quantum state tomography. However, if truncated, it can generate heuristic trial states for variational calculations, and is still subject of improvements.\cite{hejazi2023adiabatic}

Finally, it is interesting to note that the flow of information is not always from the older discipline to the newer one. Some circuit ansatz used in quantum computing can be adapted to  VMC. Inspired by the Hamiltonian variational circuit ansatz\cite{wecker2015progress} (see Appendix~\ref{app:details}) Sorella devised a method called Variational AFQM capable of obtaining state-of-the-art ground state energies of the Hubbard model for various $U/t$ parameters and dopings in the thermodynamic limit.\cite{PhysRevB.107.115133}

\subsection{Variational real-time dynamics and updates in parameters space}

Variational states are not only used for ground state calculations but they can also be used to study dynamics. The price to pay is that the variational state must be flexible enough to accurately describe also excited states, and this can be a demanding constraint, while the advantage is the ability to use much shorter circuits compared to those used, for example, for trotterization.

From a classical perspective, this area of research is very active in recent months, as it allows for countering  quantum advantage experiments in the quantum dynamics application space (cfn. Sec.~\ref{s:algo}). Obviously, the use of a variational state does not allow for exact evolution, but it is also true that the errors of a NISQ machine do not allow for it either. The balance between classical and quantum advantage for real-time dynamics will be shifted in favor of the latter when fidelity enables the simulation of sufficiently large systems for a sufficiently long time, rendering them inaccessible to classical approximation methods.\cite{daley_practical_2022}

The subfield of variational real-time dynamics also offers interesting parallels between quantum computing and (time-dependent) VMC\cite{carleo2012localization}. The formalism based on the time-dependent variational principle is the same. In practice, even the fundamental ingredients that allow for the update of variational parameters are the same: the matrix $S$ defined in Sect.~\ref{ss:optimization}. (cfn. Ref.~\cite{carleo2012localization} with Ref.\cite{li2017}). As we have seen in the case of optimization, the fact that the elements of the  $S$ matrix are subject to statistical noise is a common issue in both implementations. In this case, as well, it is reasonable to expect cross-fertilization between the two techniques, regarding both variational forms and efficient ways to evaluate the $S$ matrix.
The field of variational algorithms for real-time simulations is very active. In this area, concepts borrowed from tensor-network simulations are also useful to shorten the circuit.\cite{PhysRevResearch.5.023146,PhysRevResearch.3.033083.PRXQuantum.2.010342}

Generally speaking, variational parameters can be updated using different pseudo-dynamics $\bm{\theta}' = \bm{\theta} + \delta\bm{\theta}$ to achieve various objectives. While pure energy minimization is the most popular goal, and real-time evolution following the time-dependent variational principle is the second, there are other possibilities. Patti et al.\cite{patti2022markov} devised an iteration scheme to perform Markov chain Monte Carlo in the quantum circuit's parameter space, i.e., to sample from $p(\bm{\theta}) \sim \exp{[-\beta \langle \psi(\bm{\theta} )| H | \psi(\bm{\theta}) \rangle]}$. The resulting equation is a generalization of stochastic gradient descent that ensures detailed balance. This approach could assist in escaping local minima during VQE optimization.

Similar ideas have been proposed in the VMC context earlier.
Mazzola et. al.~\cite{2012_mazolla} showed that one can obtain an upper bound for the free energy, by sampling from $p(\bm{\theta}) \sim \sqrt{|S(\bm{\bm{\theta}})|} ~ \exp{[-\beta \langle \psi(\bm{\theta}) | H | \psi(\bm{\theta}) \rangle]}$. In VMC, this can be achieved either using a modified Langevin equation for $\bm{\theta}$ or a modified Metropolis acceptance, also known as ``penalty method''.\cite{ceperley1999penalty}

In conclusion, manipulating trial states in the presence of statistical noise is a common feature of VMC in all its formulations and scopes. Many ideas have been proposed to achieve stable parameter updates. The VQE community could profit from this established knowledge but also share its own developments and ideas to advance both fields.

\section{Classical Monte Carlo meets quantum computing}
\label{sect:sampling}

In this Section, we completely shift our perspective. Not all chemistry problems are quantum many-body ones, for example, understanding protein folding is a daunting task already in its classical force-field formulation.  Likewise, not all problems that a quantum computer can solve are genuine quantum mechanical problems. In fact, in many cases, the opposite is true: the most famous quantum algorithms, that have made the field of Quantum Information renowned, are focused on solving ``classical'' problems. For instance, Shor's algorithm provides exponential speed-up for factoring integers, and Grover's algorithm enables quadratic speed-up for searching in databases.\cite{nielsen_chuang_2010} Other examples include algorithms for linear algebra, optimization, and machine learning. Philosophically speaking, solving a purely classical problem with a quantum machine can be even more intellectually rewarding than simulating a quantum system, where the distinction between computation and simulation becomes less clear.

Up to this point, we have been exploring whether and how, well-known techniques in quantum Monte Carlo can be adapted to quantum computing to simulate many-body quantum systems. Now we are questioning the opposite: Can a quantum computer be useful in speeding up a classical Monte Carlo algorithm, where the Hamiltonian is defined solely using classical variables, e.g., classical spins? And more specifically, can we achieve this already on NISQ machines?

\subsection{Autocorrelation of a Markov chain}
\label{ss:autocorr}

\begin{figure*}[ht!]
\includegraphics[width=0.9\textwidth]{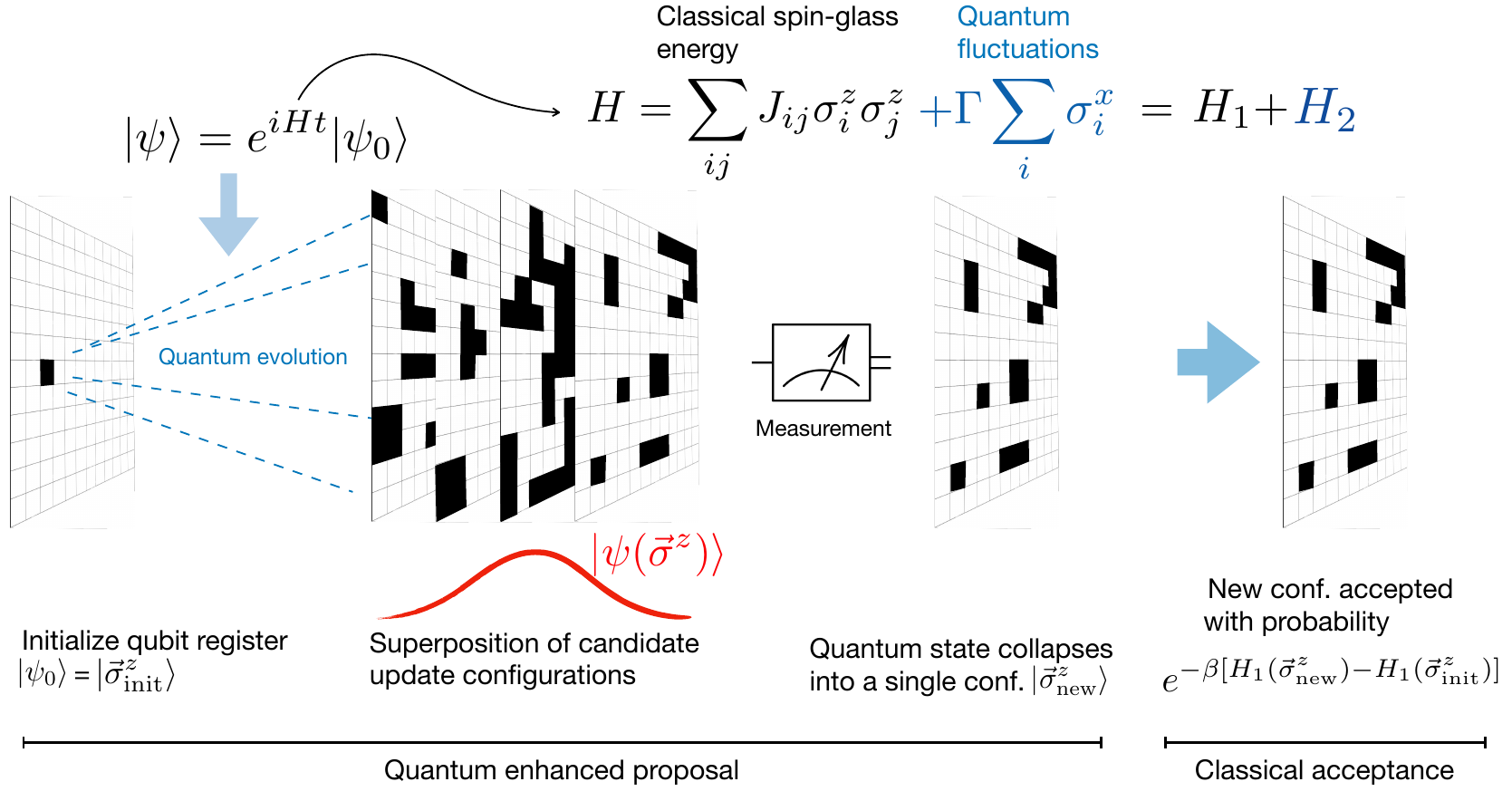}
\caption{ Schematic depiction of one quantum-enhanced Metropolis step described in Sec.~\ref{ss:enhancedMC}. The system illustrated is a 2D lattice model, with $L$ sites and a classical spin glass energy cost function, $H_1$. In this notation $|x\rangle = |\sigma^z_1,\cdots, \sigma^z_L\rangle =  |\vec{\sigma}^z \rangle$ is a bit-string, basis state of the $2^L$ dimensional Hilbert space. We start from an initial bit-string $|\vec{\sigma}^z_\textrm{init} \rangle$, that undergoes unitary evolution (in a digital quantum hardware, this can be implemented by Trotterization) using a full quantum Hamiltonian $H = H_1 + H_2$. At the end of the evolution, the measurement process collapses the time-evolved state $\ket{\psi}$ into a single bit-string $|\vec{\sigma}^z_\textrm{new} \rangle$. This concludes the proposal step $T(x,x')$. For the transverse field case, the proposal matrix is symmetric, $T(x,x') = T(x',x)$. Finally, the acceptance step is performed classically, and the new configuration may or may not be accepted. 
}
\label{fig:sketch}
\end{figure*} 

Markov chain Monte Carlo (MC) algorithms are of fundamental importance in both science and technology to understand models that lack a simple analytical solution.\cite{robert2011short,brooks2011handbook}

MC methods aim to generate statistically independent, representative configurations $x_i$, belonging to the computational space, distributed as a target  Boltzmann distribution, $\rho(x) = \exp( -\beta V(x))$, at finite inverse temperature $\beta=1/T$, and where $V(x)$ is a classical potential energy. A Markov chain MC algorithm sequentially generates these representative configurations, through a transition probability matrix $P(x,x')$, that defines which states $x,x'$ can be connected along the chain, and the relative probability of the transition $x\rightarrow x'$ (each row of the matrix $P$ is normalized to one).\cite{levin2017markov}

Among the family of Markov Chain MC algorithms, the Metropolis algorithm is certainly the most popular one. \cite{metropolis1953equation} Here the transition process takes the form $P(x,x')=T(x,x') A(x,x')$, where  $T(x,x')$ and $A(x,x')$ are, respectively, the \emph{proposal} and the \emph{acceptance} probability matrices. The algorithm works as follows: when at state $x$, a candidate trial configuration $x'$ is generated from the distribution $T(x,\cdot)$. The trial configuration is accepted with probability $A(x,x')$. If accepted, the next element of the chain becomes $x'$, otherwise, it remains $x$.
If $T$ is a symmetric matrix, then the Metropolis acceptance is given by $A(x,x') = \textrm{min}[ e^{-\beta(V(x')-V(x))},1]$.\cite{metropolis1953equation}
It is clear that the efficiency of the algorithm strongly depends on the choice of $T(x,x')$.\cite{duane1987hybrid,girolami2011riemann,wu2021unbiased}
The efficiency is given by the relaxation or mixing time, which quantifies the speed of convergence towards the equilibrium distribution $\rho(x)$, and is formally given by the inverse of the gap, $\delta$, between the largest and the second-largest (in modulus) eigenvalues of $P$.
Two limiting cases exist, the first involves a \emph{local} update scheme, based for instance on some physical intuition about the system (e.g. a single spin-flip). This usually produces a new configuration $x'$ that is similar to the parent $x$. This choice increases the acceptance rate, because $V(x)\sim V(x')$ but results also in a long sequence of statistically correlated samples, such that long simulations are needed to thoroughly explore the configuration space. On the contrary, a \emph{non-local} update scheme is more effective in producing uncorrelated samples, but usually at the expense of a vanishing acceptance rate. 

Interestingly, it took about 30 years after the invention of the Metropolis algorithm in 1953, before the introduction of efficient non-local update schemes for lattice models, the Swendsen-Wang\cite{swendsen1987nonuniversal} and the Wolff\cite{wolff1989collective} algorithms. These \emph{cluster} updates solved the \emph{critical slowing down} of MC simulations at phase transitions in ferromagnetic Ising models, but unfortunately are not as effective for frustrated models.\cite{houdayer2001cluster,barzegar2018optimization}

\subsection{Sampling from a classical Boltzmann distribution using wavefunction collapses}
\label{ss:enhancedMC}

Recently, it has been proposed the use of a quantum computer, digital or NISQ, to generate efficient non-local Metropolis updates $T(x,x')$ for spin systems.
The theoretical framework has been first introduced by Mazzola~\cite{mazzola2021sampling} in 2021.  Shortly after, Layden et. al.~\cite{layden2023quantum} demonstrated on real quantum hardware a quantum-enhanced Markov chain algorithm.

Following Ref.~\cite{mazzola2021sampling}, the general idea  is rooted in the Fokker-Plank formalism of non-equilibrium statistical mechanics in continuous systems\cite{tuanase2004metastable}. The Fokker-Plank operator $H_\textrm{FP}$ is a parent quantum Hamiltonian of the physical potential $V(x)$, which spectrum is closely connected with the number of local-minima of $V(x)$. For instance, in a double-well model, $H_\textrm{FP}$ has two lowest-lying eigenvalues and eigenstates corresponding to the symmetric(antisymmetric) combination of the two Gaussian localized states in the two wells, $|\psi_L(x)\rangle, |\psi_R(x)\rangle$.

This idea can be ported to lattice models.
Let us consider for simplicity the ferromagnetic Ising model, defined by $V=H_1$ in Eq.~\ref{eq:tfim}, as our classical potential. The task is to sample from the classical Boltzmann distribution $\exp{[-\beta H_1(x)]}$.
   Here, the autocorrelation time of a local spin-flip update scheme is dominated, at low temperatures, by the rate of the \emph{rare-event} processes that drive the system from the \emph{all-up} (left) $(\uparrow \uparrow \cdots \uparrow) := |\psi_L\rangle$ to the \emph{all-down} (right) $(\downarrow \downarrow \cdots \downarrow) :=|\psi_R\rangle$ classical states. These processes are necessarily characterized by a nucleation event, exponentially suppressed with $\beta$, and the subsequent diffusion of the domain wall separating the $\uparrow$ and the   $\downarrow$ regions\cite{du2006dynamic}.
If however, one can construct a quantum Hamiltonian $H$ such its low-lying eigenstates are mostly localized on the states  $(\uparrow \uparrow \cdots \uparrow)$ and $(\downarrow \downarrow \cdots \downarrow) $, one could then prepare and sample configurations from these states with optimal autocorrelation times, through repeated collapses of these wavefunctions.

This quantum Hamiltonian could be 
the quantum transverse field Hamiltonian in Eq.~\ref{eq:tfim}, where we add a quantum driver $H_2$ to the classical potential, $H_1$. Here, in the small $\Gamma$ limit, the two (unnormalized) lowest-lying eigenstates of $H$ are $|\psi_0 \rangle \approx |\psi_L\rangle + |\psi_R\rangle $ and $|\psi_1 \rangle \approx |\psi_L\rangle - |\psi_R\rangle $.\cite{isakov2016understanding} While the gap $E_{01}$ between these states is exponentially vanishing with the system size, the gap between the rest of the state remains $\mathcal{O}(1)$. It is clear that, by sampling configurations $x$,\footnote{here $x$ denote a lattice multi-spin configuration.} from either $|\psi_0 \rangle$ or $|\psi_1 \rangle$, we can achieve optimal autorrelation times in the large $\beta$ limit, as the states $|\psi_L\rangle$,$|\psi_R\rangle$ are sampled with equal probability.

At intermediate temperatures, the probability distribution obtained via the eigenstates projection is generally different from the classical Boltzmann distribution $\rho(x;\beta) = e^{-\beta H_1(x)}$ one aims to achieve. For this reason a standard Metropolis acceptance step needs to be performed.

One can define a valid Markov chain out of this physical intuition, rooted in (1) the preparation of a localized state, (2) a quantum propagation to prepare a linear combination of the low-energy eigenstates of $H$, and (3) a measurement to collapse into a new string state.
In Ref.~\cite{mazzola2021sampling} it is proposed to use a QPE subroutine to prepare such low-energy eigenstates.
Layden et. al.\cite{layden2023quantum} simplifies the idea and uses a Hamiltonian simulation subroutine, $e^{-i H t}$, with randomized $t$ and $\Gamma$ values at each step. Crucially they observe that such a quantum proposal update is symmetric, thus enabling a fast and practical evaluation of the acceptance step. 
The algorithm is sketched in Fig.~\ref{fig:sketch}.

A superquadratic speedup for spin glass instances is observed, i.e. polynomial speed-up of order $3.6$, compared with the best possible classical update strategy. 
Interestingly, the procedure is demonstrated on hardware where it is found that hardware errors only impact the efficiency of the chain, while the sampling remains un-biased, exactly due to the existence of a classical acceptance step after the quantum, noisy proposal move.

This quantum-enhanced Markov chain MC needs a quantum dynamics subroutine. This, in turn, can be implemented both in the fault-tolerant, the NISQ regime, and in analog simulators, provided their  architecture constraints.
Clearly, to reach a possible quantum advantage one needs to deal with the fact that, as explained in Sect.~\ref{s:hardware}, quantum gate times are much slower than classical CPUs, and this may cancel a scaling advantage for reasonable system sizes.
In particular, lattice MC simulations can be executed also on special-purpose classical hardware, such as FPGA, as demonstrated in Ref.~\cite{Janus2012}, which may enjoy an even faster logical clock speed.
Therefore, more work will be needed to assess whether this idea can bring a real benefit in this application space.

\subsection{Quantum walks and quantum Metropolis algorithms}
\label{ss:qwalks}

For the sake of completeness and clarity, it is important to mention here another family of algorithms, known as quantum walks, which share the same objective as the quantum-enhanced Markov chain algorithms described earlier: accelerating the convergence of classical Markov chains.
While the justification for quantum-enhanced Markov chain algorithm of Sect.~\ref{ss:enhancedMC} is based on physical intuition\cite{mazzola2021sampling}, and also the potential gains are assessed heuristically, quantum walks come with a more rigorous guarantee:
if they can be implemented, they provide a quadratic speed-up in autocorrelation times. This quadratic speed-up is related to concepts such as Amplitude Amplification or Grover's algorithm.\cite{grover1996fast}

There are practical and conceptual difficulties that limit the design of quantized Markov chains, particularly in the acceptance step.
In classical systems, we can always save the current configuration, previously denoted as $x$, and reuse it if the trial move leading to $x'$ is not accepted. However, in quantum computing, the \emph{no-cloning} theorem prohibits the direct copying of a quantum state.\cite{PhysRevA.61.022301}
Furthermore, the acceptance step also involves arithmetic operations that are computationally more demanding in the quantum context. It is easy to imagine that a unitary ``walk'' operator must include rotations by the following arbitrary angle
\begin{equation}
\label{eq:walkangle}
    \phi = \textrm{arcsin}\left(  \sqrt{ \mathrm{ min }[ e^{-\beta \Delta },1 ]  }   \right),
\end{equation}
where $\Delta$ is an energy difference.\cite{lemieux2020efficient}
Now, it is important to note that quantum arithmetic is much more expensive in quantum computing because it must be reversible. For instance, the most efficient way to perform the addition is still a subject of ongoing research.\cite{gidney2018halving}

The most commonly used definition of a quantum walk originates from Szegedy.\cite{szegedy:2004}
For the sake of brevity, we have to refer the reader to the comprehensive review\cite{venegas2012quantum} or to Ref.~\cite{lemieux2020efficient}, where Szegedy's algorithm is revisited from a more practical perspective, for details.
In short, for any classical Markov chain defined by the transition matrix $P(x,x')$, (see Sect.~\ref{ss:autocorr}) a quantum walk represents a quantized version of it that offers a quadratic speed-up in mixing time. 
Formally, it enhances the gap from $\delta$ to $\sqrt{\delta}$, such that the mixing time decrease quadratically from $\mathcal{O}{(1/\delta)}$ to $\mathcal{O}{(1/\sqrt{\delta})}$.

Szegedy's walk circumvents the no-cloning constraint using two copies of the graph (or lattice), and postulate the existence of a unitary operation $W$ of the form
\begin{equation}
    W \ket x \otimes \ket 0 = \sum_{x'} \sqrt{P(x,x')} \ket{x'}  \otimes \ket x .
\end{equation}
A practical implementation of the walk operator $W$ would require digital rotation of angles such as Eq.~\ref{eq:walkangle}, which are in turn evaluated by a sequence of costly arithmetic operators.
Lemieux et. al\cite{lemieux2020efficient} analyze the cost of quantum walks showing that the quadratic speed-up that they can offer is overshadowed by the cost of implementing the $W$ operator, assuming even optimistic estimates for the gate time of fault-tolerant hardware.
The quantum-enhanced method of Sec.~\ref{ss:enhancedMC} completely avoids performing the acceptance step on the quantum hardware, which is the main reason for its hardware feasibility. 

We note that it is increasingly easy to get confused with the names of the methods and the combinations of words such as ``quantum'', ``Monte Carlo'', and ``Metropolis''. In the traditional literature, as well as in this Perspective article, ``quantum Monte Carlo'' refers to the family of Monte Carlo algorithms that are executed on conventional computers but aim to solve many-body quantum problems. However, there is a community in quantum computing for which this combination of words indicates a Monte Carlo algorithm executed on quantum hardware, including Montanaro's algorithm for computing expectation values of multidimensional integrals using quantum amplitude estimation\cite{montanaro2015quantum}. This type of application finds application in finance\cite{Chakrabarti2021thresholdquantum}, and, while interesting, it is not discussed in this manuscript.

Finally, we also note that a quantum computing-based method to speed up a Monte Carlo simulation, in principle, could be used to accelerate a QMC algorithm as well. In this case, the leap would be twofold: using quantum hardware to accelerate a classical algorithm, such as a path-integral MC, to simulate quantum Hamiltonians.

Two philosophically more interesting algorithms can be mentioned for this purpose. Temme et al.\cite{temme:2011} proposed a ``quantum Metropolis algorithm'' for studying quantum many-body Hamiltonians. This method suggests performing a walk in the eigenstates of the quantum Hamiltonian, thus overcoming the sign problem.\cite{troyer_computational_2005} The algorithm includes performing and undoing QPE, an ancilla register that stores the energy, and the ability to perform on it measurements that only reveal one bit of information, enabling the acceptance step to overcome the no-cloning principle.

Yung and Aspuru-Guzik\cite{yung:2012} chose a different strategy for their ``quantum-quantum Metropolis algorithm'', finding a way to extend Szegedy's walk to quantum Hamiltonians.
The runtime performance and scaling of such approaches is still yet to be assessed.

\subsection{Sampling with quantum annealers or simulators}

The algorithms presented in Sect.\ref{ss:qwalks} require a fault-tolerant computer. However, it cannot be ruled out that an advantage in the sampling problem could come from hardwares that fall on the opposite spectrum, namely noisy quantum simulators or quantum annealers.\cite{johnson2011quantum}
First of all, let's observe that the quantum-enanched Markov chain method of Sect.~\ref{ss:enhancedMC} can be implemented not only using trotterization but also through real-time dynamics in a quantum simulator.\cite{layden2023quantum}
Furthermore, optimization and sampling tasks are closely connected.\cite{albash2018adiabatic} Special-purpose quantum simulators, called quantum annealers, have been built with the aim of optimizing large-scale spin-glass problems, but they have also been reconsidered as thermal samplers\cite{PhysRevA.94.022308} with some specific applications in machine learning.\cite{PhysRevX.7.041052, crawford2018reinforcement}

The possibility of using a quantum annealer as a sampler arises from its deviation from adiabaticity. The existence of vanishing gaps during annealing implies that at the end of the experiment, the wave function does not localize in the classical minimum of the cost function but remains delocalized, producing a distribution of read-outs. The presence of hardware noise amplifies this effect even further.

This residual distribution could resemble a thermal Boltzmann distribution of some classical Hamiltonian, close to the problem Hamiltonian originally meant to be optimized, and at some effective temperature, which is difficult to determine.\cite{PhysRevA.92.052323,PhysRevA.94.022308} However, given all the possible hardware and calibration errors, it is unlikely to hope that this approach can generate an unbiased sampling from a target distribution.

Recently, Ghamari et al.\cite{ghamari2022sampling} proposed the use of this annealing process as an importance sampler. Similarly to the quantum-enhanced Markov chain method, detailed balance is restored using a classical acceptance step. In this case, as well, a control parameter is the annealing runtime, which generates a more-or-less-localized final distribution.

Finally, Wild et. al.\cite{PhysRevLett.127.100504} proposed an
adiabatic state preparation of
Gibbs states that can also bring quantum speedup over classical Markov chain
algorithms, that could also be implemented on NISQ Rydberg atoms devices.

\section{Conclusions}
\label{s:conclu}

In this Perspective, we investigate many intersections between quantum algorithms and Monte Carlo methods.
We begin with a brief review of quantum computing applications for many-body quantum physics.
We outline the consensus that is emerging after these years in which quantum computing has become mainstream. With the availability of quantum computers with $\sim 100$ qubits and the ability to implement gate sets, albeit noisy\cite{kim2023evidence}, the field is taking on a more practical connotation beyond the traditional boundaries of quantum information theory.

We observe that different hardware platforms imply different gate frequencies, which must be taken into consideration in the perspective of achieving quantum advantage. Quantum advantage for high-accuracy many-body ground state calculations is likely to be deferred to the fault-tolerant era due to the existence of hardware noise in today's machines and the existence of highly developed classical competitors.\cite{lee2023advantagechemistry}
 Although the possibility of obtaining quantum advantage through variational methods is possible, especially in systems where classical methods struggle,\cite{wu2023variational} here we face the additional challenge posed by the presence of quantum measurements shot noise.

We then list several points of contact between quantum and classical variational methods. First, we explain the difference between the statistical noise present in conventional QMC algorithms and the noise arising from quantum measurements. Classical QMC methods feature an energy estimator - the \emph{local energy} - that enjoys the zero-variance property. This, along with stable optimizers,\cite{sorella1998green,umrigar2007alleviation} enables the optimization of wave functions featuring thousands of parameters. This is not currently possible in variational quantum computing. Even with access to an exact ground state preparation circuit, obtaining the energy with sufficient precision requires a costly number of circuit repetitions.~\cite{wecker2015progress}
It is clear that this problem arises even before the circuit optimization stage. In the current literature, this aspect is often overlooked, as several new algorithms or circuits are tested without realistic  shot noise conditions.\cite{scriva2023challenges}

We suggest that finding the quantum equivalent of the \emph{local energy} should be one priority in  variational algorithms developments. Along the same lines, we reviewed attempts and challenges in ``quantizing'' QMC methods.\cite{huggins2022unbiasing,mazzola2022exponential,zhang2022quantum,xu2022quantum}
Other areas where we expect to see cross-fertilization between the quantum and classical worlds include the development of variational forms: classically-inspired circuits for VQE, quantum-inspired ansatze for VMC. 
Several essential ingredients for variational real-time evolution and parameters optimization under noisy conditions have been already put forward in the VMC community and will be instrumental for their quantum counterparts.

Finally, after discussing how knowledge of QMC methods can provide new momentum to the development of quantum algorithms, we take the opposite direction, showing that quantum hardware can bring advantages to  Monte Carlo itself.
In this space, quantum walks have been present in the literature for several years and achieve a quadratic speed-up in autocorrelation times through the quantization of a classical Markov chain.\cite{szegedy:2004}
Their scaling is discussed, as is typical in quantum information, using an oracular form, which assumes the existence of key subroutines without delving into the details of their concrete gate-level implementation. Recently, it has been shown that these oracles require fairly long circuits. In their necessary fault-tolerant implementation, this implies absolute runtimes that are still slower than classical Monte Carlo, even for large-scale classical-spin simulations.\cite{lemieux2020efficient}

A more hardware-friendly possibility is represented by a family of methods that use a quantum computer as an importance sampler or to perform only the proposal part of a Metropolis update on the quantum hardware.\cite{mazzola2021sampling,layden2023quantum,ghamari2022sampling} Physically speaking, one simply leverages the fact that quantum measurements are uncorrelated, making them an efficient engine for sampling.
In this case, shot noise is no longer a limitation but rather becomes the computational resource for quantum advantage.\cite{mazzola2021sampling}

Overall, the purpose of this Perspective is to further connect two communities: the  quantum algorithms and Monte Carlo one. As mentioned, many methods developed in QMC can be repurposed in quantum computing. On the other hand, QMC can be a formidable competitor that can hinder or delay the quantum advantage.
This is true for quantum chemistry applications, but also optimization and beyond. 
For instance, QMC can reproduce the scaling of quantum annealing machines, for classical optimization purposes, under certain conditions.\cite{isakov2016understanding,mazzola2017tunneling,andriyash2017can}

However, the two communities can be complementary, and we hope that new impactful algorithms, either quantum or classical, will emerge thanks to this interaction, to solve important problems in chemistry and condensed matter.
\newline

\textbf{Acknowledgments.}
I acknowledge discussions on these topics in the past years with  
 S. Sorella, G. Carleo, P. Faccioli, A. Zen, K. Nakano, F. Tacchino, P. Ollitrault, S. Woerner, M. Motta, D. Layden, M. Troyer.
 I dedicate this manuscript to  the memory of Sandro Sorella, who invented many techniques mentioned above, and inspired the  whole QMC community with his creativity and enthusiasm.
I also acknowledge financial support from the Swiss 
National Science Foundation (grant PCEFP2\_203455).

\appendix

\section{Hamiltonian variational and Jastrow ansatze}
\label{app:details}

\textbf{Quantum circuit.} 
We use the Hamiltonian variational (HV) ansatz~\cite{wecker2015progress}, which is a very powerful and trasnferable ansatz, which is inspired from the adiabatic principle.
The unitary operator defining the HV ansatz is made of $d$ blocks, and each block is a product of $\ell$ operators $\hat{U}_j = \exp(\mathrm{i} \theta_j^k \hat{H}_j)$, with $j = 1, \ldots, \ell$ indexing the non-commuting terms of the Hamiltonian. For the transverse field Ising model we only need $\ell = 2$, cfn. Eq.~\ref{eq:tfim}. In this case, the full unitary operator is
\begin{equation}
\hat{U}_\text{HV}(\theta) := \prod_{i = 1}^d \hat{U}_2(\theta^i_2) \hat{U}_1(\theta^i_1),
\label{eq:hv-ansatz}
\end{equation}
which can be efficiently decomposed using one- and two-qubit quantum gates,
and the final parameterized state is
\begin{equation}
\ket{\psi(\theta)} = \hat{U}_\text{HV}(\theta) \left( \frac{\ket{0} + \ket{1}}{\sqrt{2}} \right)^{\otimes L},
\label{eq:trial-ansatz-hv}
\end{equation}
where the initial non-entangled state can be obtained from $\ket{0}^{\otimes L}$ by placing one Hadamard gate on each qubit.
The total number of parameters is $\ell d$.
In our numerical experiment, we use a state-vector emulation of the operator, based on linear algebra operations, and we do not compile our operator into a real circuit. 
Parameters are optimized using COBYLA and BFGS optimizers. These results are compatible with Ref.~\cite{wu2023variational}.

Once obtained the optimized state we emulate the noisy energy estimation using the Pauli measurement method. We sample $M_1$ spin configuration from $|\psi(\theta)|^2$ in the computational basis, and $M_2=M_1$ in the rotated basis, obtained using the $\textsf{H} \otimes \textsf{H} \cdots \textsf{H} $ operator (cfn.~\cite{peruzzo2014}).
The total cost of the energy evaluation is therefore $2 M_j$, where $M_j=M_1=M_2$ values are reported in the text.

\textbf{Classical ansatz.} For the case of VMC we use a long-range Jastrow state of the form,
\begin{equation}
\label{eq:jas}
    \psi (x) = \exp\left(  \sum_{r=1}^{L/2} \lambda_r \left( \sum_{k}^L s^z_{k} s^z_{k+r} \right) \right)
\end{equation}
with five variational parameters.
The global minimum is found when the parameters are $[0.220, 0.057, 0.030, 0.022, 0.010]$.
 This trial state yields a variational energy which is $\sim 0.1\%$ close to the exact ground state, $E_0$.
 To generate different ansatze of different quality, to showcase the zero-variance property of the local energy, we simply act on the first parameter $\lambda_1$, pulling it away from its minimum value, and towards smaller values (this is done to create a more challenging, delocalized state), while keeping the other fixed.
 When $\lambda_1 = -0.15$ the variational energy degraded up to a $10 \%$ systematic error.

 We extract $M_\textrm{VMC}$ configurations in the computational basis, since we only need this basis to compute the local energy, Eq.~\ref{eq:le_discrete} as the name suggests.
To calculate the standard deviation of the estimator we simply repeat the numerical experiment, for each ansatz and $M$ setup, 100 times.

%

\end{document}